\documentclass[preprint,aps,showpacs,prb]{revtex4}
\usepackage{amsfonts}
\usepackage{graphicx}
\usepackage{amsmath}
\usepackage{natbib}
\usepackage{epsfig}

\renewcommand\Re{\mbox{Re}\;}
\renewcommand\Im{\mbox{Im}\;}

\begin{document}

\title{Fluctuations of the Magnetization in Thin Films due to 
Conduction Electrons}

\author{A. Rebei$^{1}$}
\email{arebei@mailaps.org}

\author{M. Simionato$^{2}$}

\affiliation{ ${1.}$ Seagate Research Center, Pittsburgh, Pennsylvania 15222,USA \\
${2.}$ Department of Physics and Astronomy, University of Pittsburgh, Pennsylvania 
15260, USA }

\date{12-10-2004}

\begin{abstract}
A detailed analysis of damping and 
noise due to a {\it sd}-interaction
 in a thin ferromagnetic film 
sandwiched between two large normal 
metal layers is carried out. \ The magnetization 
is shown to obey in general a non-local
 equation of motion which differs from the 
the Gilbert equation and is extended to the 
non-adiabatic regime. \ To lowest order 
in the exchange interaction and in the limit where 
the Gilbert equation applies, 
we show that  
the  damping term is enhanced 
due to interfacial effects but it also 
shows oscillations as a function of the 
film thickness. \ The noise calculation 
is however carried out to all orders in the exchange coupling constant.
The ellipticity of the 
precession
of the magnetization is  taken into account. \ The 
damping is shown 
 to have a  Gilbert form only in the adiabatic limit while the relaxation time
 becomes 
strongly
dependent on the geometry of the thin film. \ It is also 
shown that the induced 
noise characteristic of sd-exchange is inherently colored 
in character and depends on the symmetry of the Hamiltonian 
of the magnetization in the film. \ We show that
 the sd-noise can be represented  in terms
of an
external stochastic field which is white only in the adiabatic regime. \ The 
temperature is also renormalized by the spin accumulation in the 
system. \ For large intra-atomic exchange interactions, the  
Gilbert-Brown equation is no longer valid.
\end{abstract}

\maketitle

\section{ INTRODUCTION}

\ The need for ever higher storage densities and faster retrieval 
data rates in magnetic recording is bringing out new 
fundamental physical challenges to the industry. \ In any 
physical device, the main issue is the signal to noise ratio (SNR). \ 
For nano-devices, we expect a decrease in the signal output 
and  an increase in the noise. \ Therefore any simple 
scaling-down 
of the current devices is bound to fail. \ Hence the need 
for an understanding of the noise source
  in these devices so that 
novel solutions to the SNR problem can be devised. \ In magnetic 
transition-metal based devices, the conduction electrons 
are one such source of noise. \ The stochastic Landau-Lifshitz-Gilbert 
equation, which for short we will call the Gilbert-Brown 
equation (GB)\cite{gilbert,brown},
 has 
been the main tool in studying the noise at the phenomenological 
level.\cite{smith} \ The GB equation has been quite successful 
in predicting the right form for the damping term for most of the 
measurements in this area. \ The noise term, which is represented 
by a stochastic term, is however very qualitative since any meaningful
account of the noise in a magnetic system is dependent on the 
microscopic mechanisms that give rise to the damping term in 
the first place. \ The GB equation has the simple form
\begin{equation*}
 \frac{d \mathbf{S}}{dt}=  \mathbf{S} \times \left( 
\mathbf{H}_{eff} + \alpha \frac{\mathbf{dS}}{dt} + \mathbf{h}\right) ,
\end{equation*} 
with $\mathbf{H}_{eff} $ the effective field, $\alpha$ is a damping constant 
and the stochastic field $\mathbf{h} ( t ) $ satisfies
\begin{equation*}
\langle h_i(t) h_j(t') \rangle = 2 \alpha k_B T \delta_{ij} \delta (t-t').
\end{equation*}

\ Given that more details about the physics 
of the devices are now needed to better control them, a more microscopic
treatment of the noise is in order. \ This is the subject of this 
paper. \ However, we will not be able to treat this question 
in its full generality simply because specific details about  
all systems of interest differ from one to the 
other. \ In this study, we focus mainly on a 
thin magnetic film  
geometry embedded between two normal conductors. \ Such a 
geometry happens in , e.g., a read-head  in a  
recording device and it is also 
widely used in spin-momentum transfer 
problems. \ Damping and noise due to conduction electrons 
is expected to be of importance in these systems. \ To study 
this case, we need to  compute the
 effective
action for magnons in \textit{real time} for a thin magnetic 
film embedded in a conducting paramagnet. \ We will derive and solve the 
corresponding equations of motion, then we will discuss the noise
spectrum. \ This program has already been carried out successfully 
in an 
earlier paper
\cite{rebei}, where an exactly solvable 
Caldeira-Leggett-like model has
been discussed. \ In that simple case we were able to compute 
the effective
action exactly. \ However, the model of ref. \onlinecite{rebei} is 
quite phenomenological
and not very predictive, since it involves a very large 
freedom in the
choice of the coupling constants. \ The noise expressions 
in \onlinecite{rebei} will however surface 
again in our discussion in the adiabatic limit.

\ In more realistic 
microscopic models, 
the effective action cannot be computed exactly, nevertheless it can
be computed in an approximate way.
In this paper we will discuss a very simple microscopic 
model, which 
still contains the basic physics of electrons and magnons 
in thin films
and presents quite non-trivial features. \ This model has 
been 
investigated
previously by many authors, however the relevant references 
discussing the
physics of thin films that are directly related to the work 
presented here 
are \onlinecite{berger,tser,stiles,simanek,mills,simanek2,simanek3}
. \ These latter works are primarily 
interested in effects  similar to 
the spin-momentum transfer problem of Slonczewski \cite{slon} who
studied the influence of 
a nonzero polarized current on 
 the dynamics of the 
magnetization in thin film multilayers. \ In this paper, we mainly focus 
on the single film case with and without a biased voltage. \ The case with 
a polarized current and non-collinear magnetization  will be briefly 
treated numerically in the last section.
 \ We will be mainly studying
 the finite size 
effects of the film on the dynamic of the magnetization but our method
will allow us also to address questions related to the validity
of the GB in the atomistic regime, an area of growing interest in 
recording physics.\cite{garanin,chantrell}

\ In the presence of a s-d interaction, the conduction electrons 
exercise an
effective torque on the local magnetic moment which can be put 
in a stochastic
form. \ Therefore, in addition to 
the usual
thermal magnetic noise, there will be an additional component 
due to the
conduction electrons and one of our tasks is to find out 
 when this 
contribution can be absorbed in the usual Gilbert damping term
. \ The origin of damping in ferromagnets 
is still an open problem. \ In iron, it is believed that 
the conduction electrons through the exchange interaction 
are the main channel for the dissipation. \cite{heinrich} \ In Nickel 
and Cobalt, the spin-orbit coupling is suggested to be the 
mechanism for the dissipation. \cite{korenman}$^{,}$ \cite{kunes} \ The 
calculations of \onlinecite{heinrich} and  \onlinecite{korenman} 
are however not totally self-consistent; an adjustable parameter, 
the relaxation of conduction electrons to the lattice, is
needed for a meaningful result for the 
dissipation and no treatment of noise has been 
attempted. \ Both mechanisms give a Gilbert-form 
for the damping. \ In this paper we  study damping in thin films
within the sd-model of 
ref. \onlinecite{heinrich}. \ It has been recently 
argued that the damping should reflect the geometry of 
the sample and hence the damping should have a  
non-Gilbert tensor form. \cite{safonov} \ The linear model treated in 
ref. \onlinecite{rebei} showed that the symmetry of the 
Hamiltonian has no effect on the damping. \ In this paper, we 
treat a non-linear interaction between the 
conduction electrons and the magnetization, the sd-exchange, 
 and show that in this case the damping is sensitive to the 
symmetry of the Hamiltonian  only  for high
frequencies. \ Therefore for the macroscopic average magnetization of the sample, 
the Gilbert damping is correct. \ Symmetries are important only 
for microscopic magnetization.

 \ As a result of
 the 
recent illuminating  work of Simanek \cite{simanek3}, our work will 
turn out also to be intimately related to the spin-pumping theory of 
Tserkovnyak, Brataas and Bauer (TBB) \cite{tser} that 
treated similar questions using scattering theory. \ Their damping 
is nicely expressed in terms of the mixing conductance, a quantity 
that needs to be computed by {\it ab-initio} calculations.  \ In this 
work we use a very different method which will enable us to
treat simultaneously atomic magnetic moments and macroscopic 
magnetic moments simultaneously. \ Moreover, we will be able to
give explicit expressions for the damping and noise at all frequencies 
and include 
finite size-effects of the film within the sd-model. \ Realistic
 systems 
can also be treated by this method but will  require numerical 
computations. \ Hence our results will be of interest to 
those interested in atomic simulations of magnetic 
systems, an area which is starting to become 
important for magnetic recording.\cite{chantrell}

The experimental work of Covington et al.  on
 spin momentum transfer is another motivation 
for our work.\cite{covington}  \ This latter work showed that 
in a biased spin valve with currents below the critical 
current, i.e. current needed to switch the thin layer, the resistance 
shows large 1/f-type noise in the MHz-GHz regime. \ Our system 
is similar to a spin valve except that we do not have a reference 
layer. \ This will enable us to examine 
the contribution of the sd-exchange to the 
line-width in the N/F/N structure and the spin momentum 
transfer. \ A full micromagnetic 
treatment is also given that includes the effect of non-spin flipping 
events on the spin-momentum transfer. \ Our conclusions 
will be helpful to the interpretation of the 
experiment and the micromagnetic calculations. \ We will  
show an example where the noise
 has its origins in the non-uniformity of the
in-plane magnetization.

The paper is organized as follows. In sect. II, we set up our notation and 
the Hamiltonian used in our calculations. \ We use a non-isotropic 
Hamiltonian that takes into account the ellipticity of the 
magnetization which is typical in thin magnetic 
films or local magnetic moments. \ Using 
 the real-time formalism for our model, we first
 compute the 
free propagators of the theory. \ Then,  we derive 
the effective action 
of the system by integrating out the electron degrees of 
freedom. \ We derive 
 a stochastic equation for the magnetic moment that 
is different from the  GB
equation. \ These equations 
are non-local in space and in time. \ These equations will be especially
needed in atomic-type simulations of magnetic systems where the 
local effective field is large compared to macroscopic fields. \ In sect. III, we 
discuss the limits under 
which we can recover the
 GB equation for this model in a macroscopic system N/F/N. \ We show 
that interfaces enhance the damping and the fluctuations 
of the magnetization. \ The bulk damping is assumed to be 
due to conduction electrons 
interacting with the lattice. \ In sect. IV, we show how to calculate the 
noise spectrum of the magnetization and discuss its dependencies on the anisotropy, initial conditions  and on the 
spin accumulation for the N/F/N case. \ Moreover, we show that in the  
 adiabatic limit this
sd-exchange interaction is equivalent to a stochastic 
external field with a Gaussian white noise distribution and 
effective temperature that reflects the geometry of the 
 system. \ For high frequencies, we show that 
the GB equation is no longer 
valid and that the damping reflects the symmetry of the Hamiltonian which does not 
appear in the linear or adiabatic regime. \ In sec. V, we discuss a geometry
similar to that of 
ref. \onlinecite{covington} using a macroscopic spin transfer 
model. \ Based on the quantum calculations 
in previous sections and 
the micromagnetic calculations, we suggest that the noise 
in \onlinecite{covington}
is due to thermally assisted transitions between two non-uniform 
states of the magnetization. \ Finally in the appendix, 
we give various expressions needed in the calculation of the
correlation functions of the magnetic moment and discuss the dependence 
of the damping on the symmetry of the Hamiltonian of the 
magnetic moment.

\bigskip

\section{Hamiltonian formulation and the one-loop effective action}

Let us consider a thin film of magnetic material 
interacting 
with a large external magnetic field of order one Tesla 
or more
$ \mathbf{ H}=(0,0,H) \, 
$, 
constant in time, uniform in space and directed along
 the $z$ direction which is in-plane. \ 
Let us assume the film has linear dimensions 
$D\times {L_y}\times L_z$, 
i.e. it has 
a rectangular section in the $yz$ plane (the plane parallel 
to the magnetic 
field), with area $L_y L_z$ and thickness  ${D}$  in 
the $x$ direction, with $D << L_y$ and $D << L_z$,
as shown in figure 1. 
\begin{figure}[ht]
  \begin{center}
 \mbox{\epsfig{file=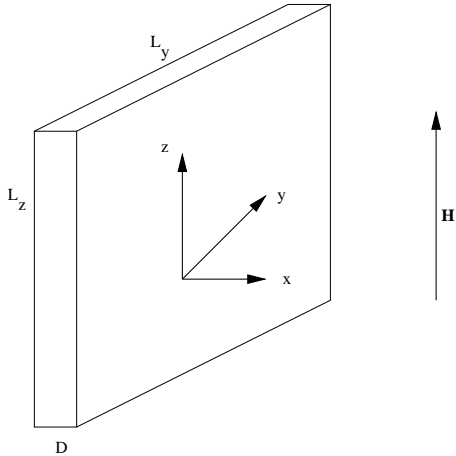,height=6 cm}}
  \end{center}
  \caption{{{The geometry of the thin film described in the text. }}}
\end{figure}
\ The magnetic film will be later 
 assumed to be sandwiched between two large 
normal conductors. \ 
We are interested in studying the effects
of the exchange 
magnetic field  due to conduction electrons 
on the average magnetization of the thin film as well as the local atomic 
moments.  Ultimately, we will 
be interested in studying
the case ${D}$ is much smaller 
than the lateral dimensions of the 
film, i.e. the thin film limit.\ The finite size effects 
related to the breaking of translation
invariance in the $x$ direction will be of primary interest to us since 
in this case the effect of the conduction electrons on the 
average magnetization is strongest. \ A path integral 
formulation proves to be very useful in problems of  
this sort. \ 

\bigskip

\begin{figure}[t]
  \begin{center}
 \mbox{\epsfig{file=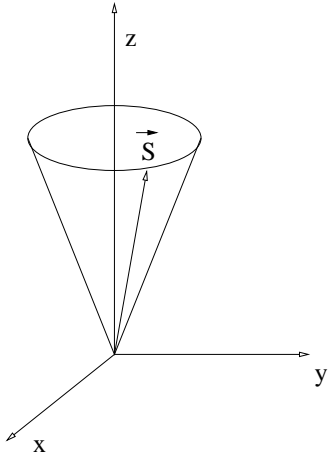,height=6 cm}}
  \end{center}
  \caption{{{The local magnetization vector. }}}
\label{vector}
\end{figure}

A ferromagnetic material  is  roughly  a
system of quasi-free electrons of spin $\mathbf{ s}$ 
(the 4s electrons) 
interacting with bound electrons of spin $\mathbf{ S}$ 
(the 3d electrons) 
via an Heisenberg Hamiltonian
\begin{equation} 
\mathcal{H}_{spin-int}=-\frac{J}{2}\int_V d^3x\;
\psi^\dagger(\mathbf x ) \vec{ \mathbf\sigma}
 \psi ( \mathbf x ) \cdot \mathbf S ( \mathbf x ),
\end{equation}
where $J$ is the interaction constant, of the order of at least 
$0.1 \; eV$ in the macroscopic case and about $10.0 \; eV$ in the microscopic 
case,  $ \vec {\mathbf \sigma} =(\sigma_1,\sigma_2,\sigma_3)$ is the vector 
with components the
Pauli matrices and $\psi$ is the 2-component 
electron field.
The 3d bound electrons are mostly aligned
with the external magnetic field and they are the source 
for the 
magnetization $\mathbf{S}$
of the material. \ The Hamiltonian of the conduction electrons 
is then given by
\begin{equation} 
\mathcal{H}_e=\int_V d^3x \;\psi^\dagger_\alpha \left( 
\mathbf x \right)\; \left[-\frac{\nabla^2}{2m}+
V (\mathbf x)\right]\psi_{\alpha} \left(
\mathbf x \right) + \mathcal {H}_{spin-int}
\end{equation}
where the potential $V(\mathbf x)$ is a spin independent confining 
potential of the structure. \ For a large 
external field $H$, the $S^z$-component of the magnetization 
can be taken 
to be a constant and hence its interaction can  be absorbed 
in the diagonal part of the energy.

\ The simplest possible effective Hamiltonian we can construct 
for the magnetization is
\begin{equation}
\mathcal{H}_{spin}=\sum_c
\mathcal{H}_{spin,c}= \int_V d^3x\; \left[
 - \mathbf H\cdot\mathbf S+
\frac12 c_{xx} S^x(\mathbf x)^2+\frac12  c_{yy} S^y(\mathbf x)^2\right]\;.
\end{equation}
where $c$ is the cell index with volume $v=V/N$. \ In micromagnetic simulations, $v$ is of the  order of 
$10 \; nm^3$ while in atomistic calculations it is of the order of the $0.4 \; nm^3$.  \ We have 
 neglected higher powers of the $S^x_c$ and $S^y_c$ components.
This can be justified once 
we notice that in a magnetic material the spins are mostly oriented in the 
direction of the magnetic field, i.e. the spin vector has locally 
$x$ and $y$ 
components which are small with respect to the $z$ component:
\begin{equation}
\mathbf S_c = (S_c^x, S_c^y, S_c^z),
\quad |S_c^x|,|S_c^y|\ll S_c^z
\end{equation} 
In other words, in this
paper we will consider the case in which the spin vector has a small
angle with the $z$ axis (see Figure \ref{vector}). \ Micromagnetic calculations
show that an angle of $20$ degrees can still be considered small
.\  The dynamics of large angles
and in particular the possibility of magnetization switching
 is quite interesting
too, but it cannot be addressed within the 
approximations used here. \cite{rebei_ox}

\begin{figure}[ht]
  \begin{center}
 \mbox{\epsfig{file=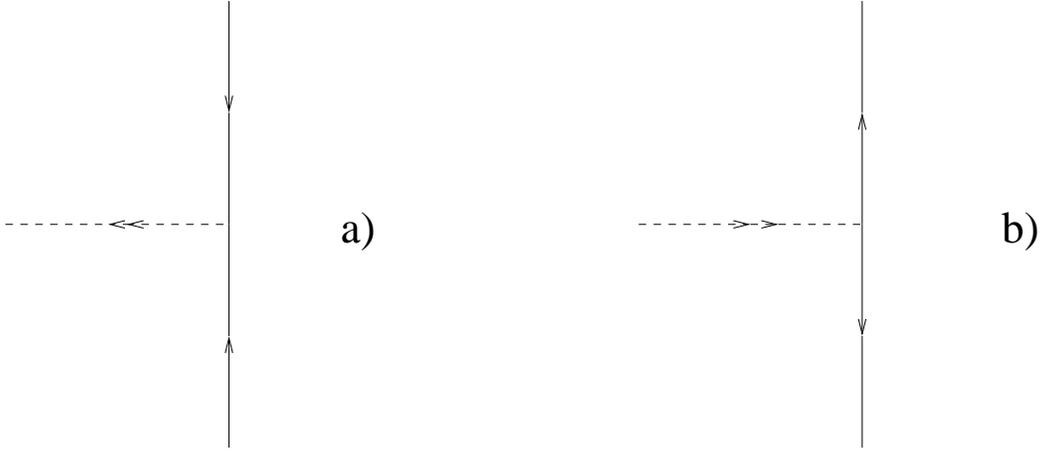,height=6 cm}}
  \end{center}
  \caption{{{\small Conservation of angular momentum in the $z$ direction}}}
\label{mom}
\end{figure}

The advantage of the small angle approximation, i.e. 
taking $S_c^z$ time-independent, is the simplification 
of the commutation relations, 
 $\mathbf{S} \times \mathbf{S}=i \mathbf{S}$, \cite{kittel}$^{,}$
\cite{kittel2} 
which become
\begin{equation}\label{ccc}
[S^x_c,S^y_{c'}]=i S^z_c\delta_{cc'}\;.
\end{equation}
This implies that $S^y_c$ is canonically conjugate 
to $S_c^x$. \ We neglect any exchange stiffness between the 
cells.  
\ Next, we consider  the case of an infinite 
wavelength spin wave 
with wave-vector $k=0$. \ This 
corresponds to an homogeneous spin field.
The Heisenberg equation of motion can be derived from 
the Hamiltonian and
the commutation relation and they assume the form
\begin{equation}\label{S.eq}
\dot S_x= c_{yy} S_y\;,\quad \dot S_y=-c_{xx} S_x\;.
\end{equation}
The coefficients $c_{xx}$ and $c_{yy}$ are related to 
the anisotropies of
the medium and we are especially interested in the 
strongly anisotropic
case, $c_{xx}\gg c_{yy}$.
The Heisenberg equations 
of motion  can be trivially solved in the form
$$
\begin{pmatrix} S_x(t)\cr S_y(t)\end{pmatrix}=e^{\mathcal M(t-t_0)}
\begin{pmatrix} S_x(t)\cr S_y(t)\end{pmatrix},
$$
where $\mathcal M$ is the matrix
\begin{equation}\label{def.M}
\mathcal M=\begin{pmatrix} 0&c_{yy}\cr-c_{xx}&0\end{pmatrix}.
\end{equation}
More explicitly, we have
\begin{eqnarray*}
S_x(t)&=&\cos\omega_0(t-t_0)\; S_x(t_0)+\sqrt{\frac{ c_{yy}}{c_{xx}}}\;
\sin\omega_0(t-t_0)\; S_y(t_0),
\\ 
S_y(t)&=&-\sqrt{\frac{ c_{xx}}{c_{yy}}}\;\sin\omega_0(t-t_0)\; S_x(t_0)+ 
\cos\omega_0(t-t_0)\; S_y(t_0)\; ,
\end{eqnarray*}
where $
\omega_0=\sqrt{c_{xx} c_{yy}}$ is the frequency of the elliptic
precession of the magnetization. \ In the rest of the paper, we 
 calculate the effect of the conduction electrons on this 
solution in both low and high frequency limits.
\ Since the interaction with 
the non-dynamical $S^z$ component has been 
already accounted for in the 
Hamiltonian of the conduction electrons, 
only the $\sigma_1$ and $\sigma_2$ terms enter in the interaction
with the dynamical spin field. In momentum 
space this interaction term reads
\begin{equation} 
\mathcal{H}_{int}=-\frac{J}{2 N^{1/2}} \sum_{\mathbf{qk}} \left[
a_{\mathbf{q}-\mathbf{k}}^\dagger \sigma_1 a_\mathbf{q} S^x_\mathbf{k}
+a_{\mathbf{q}-\mathbf{k}} ^\dagger\sigma_2 a_q S^y_\mathbf{k}\right]\; .
\end{equation}
\ The interaction 
Hamiltonian $\mathcal{H}_{int}$
commutes with the total Hamiltonian. \ This means
that only two interactions are possible: a) a spin up 
electron
$+1/2$ emits a $+1$ magnon and becomes a spin 
down $-1/2$ electron; 
b) a spin down $-1/2$ electron absorbs a spin $+1$ 
magnon and becomes 
a spin up $+1/2$ electron.  These interactions are 
represented in figure \ref{mom}.

\ Next, we calculate the effective action of the theory to 
first order in $J^2$. \ This is done by eliminating the electronic 
degrees of freedom after which, we get 
a stochastic equation for the 
magnetization valid in the non-adiabatic regime.

First we compute the
 free propagators which are obtained by inverting the 
diagonal differential
operators $i\partial_t-\mathcal E_k$ for the electron field and
the differential operator
\begin{equation}\label{D^-1}
\mathcal{D}^{-1}_{ij}=\left.\frac{\delta \mathcal S}
{\delta S_{-k}^i(t)\delta S_k^j(t')}\right|_{\phi=0}=
-\begin{pmatrix}c_{xx}&\partial_t\cr -\partial_t& c_{yy}
\end{pmatrix}\delta_P(t-t')
\end{equation}
for the spin field. \ Here $\mathcal E_k$ and $\mathcal S$ are 
the diagonal parts of the action for the electrons and 
the  magnetization, respectively. \ Therefore, one has 
to solve the differential equations for the corresponding
propagators $\mathcal G_P$ and $\mathcal D_P$:

\begin{equation}\label{G.eq}
(i\partial_t-\mathcal E_k)\mathcal{G}_P(t,t',k)=\delta_P(t-t')
\end{equation}
and
\begin{equation}\label{D.eq}
\mathcal{D}^{-1} \mathcal{D}_P(t,t',k)=\delta_P(t-t')
\end{equation}
where the delta function is defined on the 
closed-time path (CTP) $P$  and the boundary
conditions are such that the spin field is periodic 
on the path P while the electron field is anti-periodic.\cite{CTP}
Notice that eq. \ref{G.eq}
is a $2\times2$ matrix in the spin space whereas eq. \ref{D.eq}
is a $2\times2$ matrix in the complex plane. The first equation is
diagonal and can easily be solved with solution
$$
\mathcal{G}_P(t-t')=-ie^{-i\omega_k(t-t')}\Theta_P(t-t')+A e^{-i\omega_k(t-t')}
$$
where the matrix $A$ is the integration constant corresponding 
to a generic solution of the homogeneous equation. 
The boundary condition $\mathcal{G}_P(t_0,t')=-
e^{\beta\mu}\mathcal{G}_P(t_0-i\beta,t')$ fixes $A=i f_k$. Therefore
\begin{equation}\label{G.prop}
\mathcal{G}_P(t,t',\mathbf k)=-i\left(\Theta_P(t-t')-f_k\right)e^{-i\mathcal E_k (t-t')}\;.
\end{equation}
The second equation \ref{D.eq} seems more complicated, but actually
can be simplified by multiplying both sides by $i\sigma_2$ and using the
identity $
i\sigma_2 \mathcal{D}^{-1}=\partial_t-\mathcal M\;,
$
where $\mathcal M$ satisfies $
\mathcal M^T=-\sigma_2\mathcal M \sigma_2\;$. \ A simple 
computation gives
\begin{equation}\label{D.prop}
\mathcal{D}_P(t,t')=e^{\mathcal M (t-t_0)}\;i\sigma_2\left[\Theta_P(t-t')
+n(i\mathcal M^T)\right]e^{\mathcal M^T (t'-t_0)}\;,
\end{equation}
where the integration constant matrix 
$n(i\mathcal M)$ is fixed by the 
boundary conditions $\mathcal{D}_P(t_0,t') =
\mathcal{D}_P(t_0-i\beta,t')$ to be
\begin{equation}
n(i\mathcal M^T)=\frac1{\exp(i\beta \mathcal M^T)-1}\;.
\end{equation}
Notice that the time evolution is trivial and an 
explicit computation
gives
\begin{eqnarray}
\mathcal{D}_P(t,t')=\begin{pmatrix}
-\frac{c_{yy}}{\omega_0}\sin\omega_0(t-t')&\cos\omega_0(t-t')\cr
-\cos\omega_0(t-t')&-\frac{c_{xx}}{\omega_0}\sin\omega_0(t-t')
\end{pmatrix}
\left[\Theta_P(t-t')+n(i\mathcal M^T)\right]
\end{eqnarray}
Since $\mathcal M$ has eigenvalues $+i\omega_0$ and $-i\omega_0$,
the eigenvalues of $n(i\mathcal M^T)$ are $n(-\omega_0)$ and   $n(\omega_0)$
respectively, i.e., they are regular Bose-Einstein distributions. 
A little algebra allows to derive the free spectrum as
$$
<:\{S^i(t),S^j(t)\}:>=\begin{pmatrix}\frac{c_{yy}}{\omega_0}
\frac{\sinh\beta\omega_0}{1-\cosh\beta\omega_0}&-i\cr
i&\frac{c_{xx}}{\omega_0}
\frac{\sinh\beta\omega_0}{1-\cosh\beta\omega_0}
\end{pmatrix}\;.
$$

In the 
isotropic case, things are simpler since $\mathcal M$ is proportional to
$i\sigma_2$ and therefore it commutes with $i\sigma_2$. 
\ We also observe that in the CTP formalism, 
 each of the matrix equations \ref{G.prop} and 
\ref{D.prop}, 
corresponds 
to an additional $2\times2$ matrix of equations in the
Schwinger formalism, depending on the position of $t$ and $t'$ in
the path $P_1$ directed along the positive evolution in time
and $P_2$ directed in the opposite direction.\cite{schwinger} In particular

\begin{eqnarray}
\mathcal{G}_P(t,t')=G_\mathbf{k}^>(t,t')=&-i (1-f_\mathbf{k}) e^{-i\varepsilon_\mathbf{k} (t-t')}\;,
& t\in P_2,\; t'\in P_1\\
\mathcal{G}_P(t,t')=G_\mathbf{k}^<(t,t')=&i f_\mathbf{k} e^{-i\varepsilon_\mathbf{k} (t-t')}\;,
& t\in P_1,\; t'\in P_2
\end{eqnarray}
It is easy to check that the previous
propagators are consistent with the operator's expressions

\begin{equation}
G_\mathbf{k}^>(t,t')=-i<a_\mathbf{k}(t)a_\mathbf{k}^\dagger(t')>_c,\quad
G_\mathbf{k}^<(t,t')=i<a_\mathbf{k}^\dagger(t')a_\mathbf{k}(t)>_c
\end{equation}
\begin{equation}
D_\mathbf{k}^{ij>}(t,t')=-i<S^i_{-\mathbf k}(t)S^j_\mathbf k (t')>_c,\quad
D_\mathbf{k}^{ij<}(t,t')=-i<S^j_\mathbf{k}(t')S^i_{-\mathbf{k}}
(t)>_c\;.
\end{equation}

\bigskip

\ Next, we include the effect of the conduction electrons on the 
magnetization. \ Using the functional formulation,\cite{schwinger}
 it is clear how to extract the effect
of the electrons on the local magnetization field: it is enough to
integrate out the fermionic degrees of freedom and to compute the
effective action for the magnetization. Since the original action is
quadratic in 
$\psi$ and $\psi^\dagger$, the functional integral can be performed.  \ If we use a coherent state representation for the 
magnetization and the conduction electrons,
the generating functional
 for the problem has the form \cite{orland} 
\begin{equation}
Z\left[ J\right] =\int dz_{1}dz_{1}^{\ast}\int d\zeta_{1,i}d\zeta
_{1,i}^{\ast}\exp\left[ -z_{1}z_{1}^{\ast}-\zeta_{1,i}\zeta_{1,i}^{\ast }%
\right] \left\langle z_{1},-\zeta_{1}\left| \rho U^{\dagger}\left(
J_{2}\right) U\left( J_{1}\right) \right| z_{1},\zeta_{1}\right\rangle
\end{equation}
where $z$'s represent the d-electrons degrees of freedom while $\zeta$'s represent
 the conduction electrons. $J_1$ and $J_2$ are the usual virtual 
sources. \ The propagator inside is given by 
\begin{align}
\left\langle \left| {}\right| \right\rangle & =\int dzdz^{\ast}\int d\zeta
d\zeta^{\ast}\exp\left[ z_{1}^{\ast}z\left( t_{0}-i\beta\right)
+\zeta_{1,kj}^{\ast}\zeta_{kj}\left( t_{0}-i\beta\right) \right. \\
& \left. +i\int_{t_{0}}^{t_{0}-i\beta}ds\left(
iz^{\ast}\partial_{s}z-H^{J}\left( z^{\ast},z\right) \right) \right.  \notag
\\
& \left. +i\int_{t_{0}}^{t_{0}-i\beta}ds\left(
i\zeta_{kj}^{\ast}\partial_{s}\zeta_{kj}+\zeta_{kj}^{\ast}\mathcal{G}%
_{kjj^{^{\prime}}}^{-1}\zeta_{kj^{^{\prime}}}\right) \right]  \notag
\end{align}
where 
\begin{equation}
\mathcal{G}_{kjj^{^{\prime}}}^{-1}\left( \mathbf k,\mathbf p\right) =\left[ 
\begin{array}{cc}
\left( i\partial_{s}-\varepsilon_{1}\left( \mathbf k\right) \right) \delta\left(
\mathbf k - \mathbf p\right) & \frac{J}{2}z\left( \mathbf k - \mathbf 
p\right) \\ 
\frac{J}{2}z^{\ast}\left( \mathbf k - \mathbf p\right) & \left( i\partial_{s}-\varepsilon_{2}\left(
\mathbf k\right) \right) \delta\left( \mathbf k - \mathbf p\right)
\end{array}
\right]
\end{equation}
the variable $s$ represents the time along the CTP $P$. \ In 
the above we have set the total number of magnetic cells $N = 1$ since we the same 
discussion applies to a local atomic moment. \
Next we
integrate out the conduction electrons degrees of freedom. \  Using

\begin{equation}
\int d\zeta^{\ast}d\zeta\exp\left[ i\int ds\zeta^{\ast}\mathcal{G}^{-1}\zeta%
\right] =\exp\left[ -Tr\ln\mathcal{G}^{-1}\right],
\end{equation}
and expanding the logarithmic term gives exchange terms of all orders in the
coupling constant $J$. In the following, we keep only quadratic terms. The
conduction electron propagator satisfies the equation 
\begin{equation}
\left[ \left( i\partial_{s}-\varepsilon_{i}\left( \mathbf k\right) \right)
\delta_{ij}\delta_{k,k^{^{\prime}}}-V_{ij}\left( \mathbf k - \mathbf k ^{^{\prime}}\right) %
\right] \mathcal{G}_{jp}^{kk^{^{\prime}}}\left( s-s^{^{\prime}}\right)
=-\delta_{ip}\left( s-s^{^{\prime}}\right) \delta\left(
\mathbf k - \mathbf k ^{^{\prime}}\right)
\end{equation}
with $V$, the interaction term is given by 
\begin{equation}
V\left( \mathbf k - \mathbf k ^{^{\prime}}\right) =\frac{J}{2}\left[ 
\begin{array}{cc}
0 & z\left( \mathbf k - \mathbf k ^{^{\prime}}\right) \\ 
z^{\ast}\left( \mathbf k - \mathbf k ^{^{\prime}}\right) & 0
\end{array}
\right].
\end{equation}
This equation is solved by iteration, assuming that higher order terms are
 small. \ The large Zeeman-type term has been included in the $\varepsilon$ 
term. \ The
propagator $\mathcal{G}$ is then given in terms of the propagator $\mathcal{G%
}^{(0)}$ which is a solution of the following equation

\bigskip

\begin{equation}
\left( i\partial _{s}-\varepsilon _{i}\left( k\right) \right) \mathcal{G}%
_{ij}^{(0)k,k^{^{\prime }}}\left( s,s^{^{\prime }}\right) =-\delta
_{ij}\left( s-s^{^{\prime }}\right) \delta \left( \mathbf{k}
-\mathbf{k}^{^{\prime }}\right)
\end{equation}
If we define the functions $G^{<}$ and $G^{>}$ to be 
\begin{align}
G^{<}\left( s,s^{^{\prime }}\right) & =-i f(\varepsilon)\exp \left[
-i\varepsilon \left( s-s^{^{\prime }}\right) \right] ,  \notag \\
G^{>}\left( s,s^{^{\prime }}\right) & =i\left( 1-f(\varepsilon)\right) \exp 
\left[ -i\varepsilon \left( s-s^{^{\prime }}\right) \right]
\end{align}
then formally, we have 
\begin{equation}
\mathcal{G}^{\left( 0\right) }\left( s,s^{^{\prime }}\right) =G^{<}\left(
s,s^{^{\prime }}\right) \Theta _{P}\left( s^{^{\prime }}-s\right)
+G^{>}\left( s,s^{^{\prime }}\right) \Theta _{P}\left( s-s^{^{\prime
}}\right),
\end{equation}
where $\Theta _{P}\left( s-s^{^{\prime }}\right) $ is the step function
along the closed-time path $P$. Hence, the presence of conduction electrons in the
thin film will give rise to an effective action on the
d-electrons. \ To order $J^2$ , the logarithmic term becomes 

\begin{align}
Tr\ln \mathcal{G}& =\frac{1}{2}\sum_{k,k^{^{\prime }}}J^{2}\int
dsds^{^{\prime }}\mathcal{G}_{ii,kk}^{(0)}\left( s,s^{^{\prime }}\right)
V_{ij,kk^{^{\prime }}}\left( s^{^{\prime }}\right) \mathcal{G}%
_{jj,k^{^{\prime }}k^{^{\prime }}}^{(0)}\left( s^{^{\prime }},s\right)
V_{ji,k^{^{\prime }}k}\left( s\right) \\
& =J^{2}\sum_{k,k^{^{\prime }}}\int_{c}dsds^{^{\prime }}z\left( s^{^{\prime
}},\mathbf{k}-\mathbf{k}^{^{\prime }}\right) z^{\ast }\left( s,\mathbf{k}^{^{\prime }} - \mathbf{k}\right) \exp 
\left[ -i\left( \varepsilon _{1}\left( \mathbf{k}\right) -\varepsilon _{2}\left(\mathbf
k^{^{\prime }}\right) \right) \left( s-s^{^{\prime }}\right) \right]  \notag
\\
& \times \left[ \Theta \left( s-s^{^{\prime }}\right) f\left( \varepsilon
_{2}\left( \mathbf k^{^{\prime }}\right) \right) \left( 1-f\left( \varepsilon
_{1}\left( \mathbf k\right) \right) \right) +\Theta \left( s^{^{\prime }}-s\right)
f\left( \varepsilon _{1}\left( \mathbf k \right) \right) \left( 1-f\left( \varepsilon
_{2}\left( \mathbf k^{^{\prime }}\right) \right) \right) \right] ,  \notag
\end{align}
which can be written in a more compact form as follows 
\begin{align}
Tr\ln \mathcal{G}& =J^{2}\sum_{k,k^{^{\prime
}}}\int_{t_{0}}^{t}dtdt^{^{\prime }}\exp \left[ -i\left( \varepsilon
_{1}\left( \mathbf k\right) -\varepsilon _{2}\left( \mathbf k^{^{\prime }}\right) \right)
\left( t-t^{{\prime }}\right) \right] \\
& \times z_{i}^{\ast }\left( t,\mathbf k^{^{\prime }}-\mathbf k\right) \mathfrak{A}
_{ij} \left( t-t^{^{\prime }},\mathbf k,\mathbf k^{^{\prime }}\right)
z_{j}\left( t^{^{\prime }},\mathbf k - \mathbf k^{^{\prime }}\right)  \notag
\end{align}
In matrix form, 
\begin{eqnarray}
\mathfrak{A}_{11} \left( t-t^{^{\prime }},\mathbf k,\mathbf k^{^{\prime }}\right)
&=&\Theta \left( t-t^{^{\prime }}\right) f\left( \varepsilon _{2}\left(
\mathbf k^{^{\prime }}\right) \right) \left( 1-f\left( \varepsilon _{1}\left(
\mathbf k\right) \right) \right) \nonumber \\ 
&& +\Theta \left( t^{^{\prime }}-t\right) f\left(
\varepsilon _{1}\left( \mathbf k\right) \right) \left( 1-f\left( \varepsilon
_{2}\left( \mathbf k^{^{\prime }}\right) \right) \right), \\
\mathfrak{A}_{12} \left( t-t^{^{\prime }},\mathbf k,\mathbf k^{^{\prime }}\right)
&=&f\left( \varepsilon _{1}\left( \mathbf k\right) \right) \left( 1-f\left(
\varepsilon _{2}\left( \mathbf k^{^{\prime }}\right) \right) 
\right) ,\\
\mathfrak{A}_{21} \left( t-t^{^{\prime }},\mathbf k,\mathbf k^{^{\prime }}\right)
&=&f\left( \varepsilon _{2}\left( \mathbf k^{^{\prime }}\right) \right) \left(
1-f\left( \varepsilon _{1}\left( \mathbf k\right) \right) \right), \\
\mathfrak{A}_{22} \left( t-t^{^{\prime }},\mathbf k,\mathbf k^{^{\prime }}\right)
&=&\Theta \left( t^{^{\prime }}-t\right) f\left( \varepsilon _{2}\left(
\mathbf k^{^{\prime }}\right) \right) \left( 1-f\left( \varepsilon _{1}\left(
\mathbf k\right) \right) \right) \nonumber \\ 
&& +\Theta \left( t-t^{^{\prime }}\right) f\left(
\varepsilon _{1}\left( \mathbf k\right) \right) \left( 1-f\left( \varepsilon
_{2}\left( \mathbf k^{^{\prime }}\right) \right) \right).
\end{eqnarray}
where $f\left( \varepsilon_\sigma \right)$ is the Fermi-Dirac distribution
for spin-up $(\sigma=1)$ and spin down $(\sigma=2)$, respectively.
The effective action of the magnetization is now given by 
\begin{align}
S^{eff}& =\int dt\left[ i z_{1}^{\ast }\partial _{t}z_{1}-H^{J}\left(
z_{1}^{\ast },z_{1}\right) \right] \\
& -\int dt\left[ iz_{2}^{\ast }\partial _{t}z_{2}-H^{J}\left( z_{2}^{\ast
},z_{2}\right) \right] +iTr\ln \mathcal{G}  \notag
\end{align}
where $z =S_{+}=S_{x}\pm iS_{y} $.  Next, we make a change of variables 
\begin{equation}
\left( 
\begin{array}{c}
z_{1} \\ 
z_{2}
\end{array}
\right) =\left( 
\begin{array}{cc}
1 & 1/2 \\ 
1 & -1/2
\end{array}
\right) \left( 
\begin{array}{c}
\mathfrak{Z} \\ 
\mathfrak{z}
\end{array}
\right) \equiv { T}\left( 
\begin{array}{c}
\mathfrak{Z} \\ 
\mathfrak{z}
\end{array}
\right)
\end{equation}
The new quadratic form is then 
\begin{equation}
\sum_{ij}z_{i}^{\ast }\mathfrak{A}_{ij}z_{j}=\left( 
\begin{array}{cc}
\mathfrak{Z}^{\ast } & \mathfrak{z}^{\ast }
\end{array}
\right) { T}^{T}\mathfrak{A}{ T}\left( 
\begin{array}{c}
\mathfrak{Z} \\ 
\mathfrak{z}
\end{array}
\right) .
\end{equation}
This form first appeared in Schwinger's paper \cite{schwinger} but 
is often cited under the name of the Keldysh form \cite{keldysh} , 
\begin{align}
\mathcal{U}\left( t-t^{^{\prime }},\mathbf k,\mathbf k^{^{\prime }}\right) & ={ T}^{T}%
\mathfrak{A}{ T}, \\
\mathcal{U}_{11}\left( t-t^{^{\prime }},\mathbf k,\mathbf k^{^{\prime }}\right) & =0 \\
\mathcal{U}_{12}\left( t-t^{^{\prime }},\mathbf k,\mathbf k^{^{\prime }}\right) & =\Theta
\left( t^{^{\prime }}-t\right) \left[ f\left( \varepsilon _{1}\left(
\mathbf k\right) \right) -f\left( \varepsilon _{2}\left( \mathbf k^{^{\prime }}\right)
\right) \right] ,\\
\mathcal{U}_{21}\left( t-t^{^{\prime }},\mathbf k,\mathbf k^{^{\prime }}\right) & =-\Theta
\left( t-t^{^{\prime }}\right) \left[ f\left( \varepsilon _{1}\left(
\mathbf k\right) \right) -f\left( \varepsilon _{2}\left( \mathbf k^{^{\prime }}\right)
\right) \right], \\
\mathcal{U}_{22}\left( t-t^{^{\prime }},\mathbf k,\mathbf k^{^{\prime }}\right) & =\frac{1}{2}%
f\left( \varepsilon _{1}\left( \mathbf k\right) \right) \left( 1-f\left( \varepsilon
_{2}\left( \mathbf k^{^{\prime }}\right) \right) \right) +\frac{1}{2}f\left(
\varepsilon _{2}\left( \mathbf k^{^{\prime }}\right) \right) \left( 1-f\left(
\varepsilon _{1}\left( \mathbf k\right) \right) \right).
\end{align}
It is easy to see from these definitions of the kernels $\mathcal{U}%
_{ij}$ that
\begin{align}
\mathcal{U}_{12}\left(  \mathbf{k},\mathbf{k}^{\prime},t\right)    &
=-\mathcal{U}_{21}\left(  \mathbf{k},\mathbf{k}^{\prime},-t\right)  ,\\
\mathcal{U}_{22}\left(  \mathbf{k},\mathbf{k}^{\prime},t\right)    &
=\mathcal{U}_{22}\left(  \mathbf{k},\mathbf{k}^{\prime},-t\right)  .
\end{align}
\ We can now introduce a new field $\eta $ into the theory. This will mimic a
random Gaussian field in the semi-classical equations. In the generating
functional $Z\left[ J\right] $, we replace the quadratic term in $\mathfrak{z}$
and $\mathfrak{z}^{\ast }$ by linear terms, 
\begin{align}
& \exp \left[ -\left(\frac{J}{2}\right)^{2}\sum_{kk^{^{\prime }}}\int dt\int dt^{^{\prime }}\mathfrak{z}%
^{\ast }\left( t,\mathbf k^{^{\prime }}-\mathbf k\right) \mathcal{U}_{22}\left( t-t^{^{\prime
}},\mathbf k,\mathbf k^{^{\prime }}\right) \right.  \notag \\
& \left. \qquad \exp \left[ -i\left( \varepsilon _{1}\left(
\mathbf k\right) -\varepsilon _{2}\left(\mathbf  k^{^{\prime }}\right) \right) \left(
t-t^{^{\prime }}\right) \right] \mathfrak{z}\left( t^{^{\prime }},\mathbf  k- \mathbf k^{^{\prime
}}\right) \right] \notag \\
& =\int d\eta ^{\ast }d\eta \exp \left[ -\sum_{kk^{^{\prime }}}\int dt\int
dt^{^{\prime }}\eta ^{\ast }\left( t,\mathbf  k\right) \mathfrak{D}^{-1}\left(
t, \mathbf k;t^{^{\prime }},k^{^{\prime }}\right) \eta \left( t^{^{\prime
}},\mathbf  k^{^{\prime }}\right) \right.  \notag \\
& \left. \qquad +i\sum_{k}\int dt\eta ^{\ast }\left( t,\mathbf  k\right) \mathfrak{z}%
\left( t,\mathbf  k\right) +\mathfrak{z}^{\ast }(t,\mathbf  k)\eta (t,\mathbf  k)\right]  \; ,
\end{align}
where the kernel $\mathfrak{D}$ is given by
\begin{eqnarray}
\mathfrak{D}\left( t, \mathbf k;t^{^{\prime }},\mathbf  k^{^{\prime }}\right) &=& \frac{J^{2}}{2}\exp \left[
-i\left( \varepsilon _{1}\left(\mathbf  k\right) -\varepsilon _{2}\left(\mathbf  k^{^{\prime
}}\right) \right) \left( t-t^{^{\prime }}\right) \right] \\
&&\times \left( f\left( \varepsilon _{1}\left(\mathbf  k\right) \right) -f\left(
\varepsilon _{2}\left(\mathbf  k^{^{\prime }}\right) \right) \right) \coth \left[ 
\frac{\beta \left( \varepsilon _{1}\left(\mathbf k\right) -\varepsilon _{2}\left(\mathbf
k^{^{\prime }}\right) \right) }{2}\right]. \notag
\end{eqnarray}
In the bulk, this noise kernel will in general depend on the relaxation 
of the conduction electrons due to phonons which will appear in the 
exponential term on the right hand side.
After this integral transformation, the effective action becomes 
\begin{align}
iS^{eff}& =-\int dt\left\{ \frac{1}{2}\mathfrak{Z}^{\ast }\partial _{t}\mathfrak{z}+%
\frac{1}{2}\mathfrak{z}^{\ast }\partial _{t}\mathfrak{Z}-\frac{1}{2}\mathfrak{Z}\partial
_{t}\mathfrak{z}^{\ast }-\frac{1}{2}\mathfrak{z}\partial _{t}\mathfrak{Z}^{\ast }\right| \nonumber
\\
& \left. -i\Omega \left( \mathfrak{Z}^{\ast }\mathfrak{z}+\mathfrak{z}^{\ast }\mathfrak{Z}%
\right) -iK\left( \mathfrak{Z}^{\ast }\mathfrak{z}^{\ast }+\mathfrak{zZ}\right) \right\}
\notag \\
& + \left(\frac{J}{2}\right)^{2}\sum_{kk^{^{\prime }}}\int dt\int dt^{^{\prime }}\exp \left[
-i\left( \varepsilon _{1}\left(\mathbf  k\right) -\varepsilon _{2}\left(\mathbf  k^{^{\prime
}}\right) \right) \left( t-t^{^{\prime }}\right) \right]  \notag \\
& \times \left\{ \mathfrak{Z}^{\ast }\left( t,\mathbf k^{^{\prime }}-\mathbf k\right) \mathcal{U}%
_{12}\left( t-t^{^{\prime }},\mathbf k,\mathbf k^{^{\prime }}\right) \mathfrak{z}\left(
t^{^{\prime }},\mathbf k-\mathbf k^{^{\prime }}\right) \right.  \notag \\
& \left. +\mathfrak{z}\left( t,\mathbf k^{^{\prime }}-\mathbf k\right) \mathcal{U}_{21}\left(
t-t^{^{\prime }},\mathbf k,\mathbf k^{^{\prime }}\right) \mathfrak{Z}\left( t^{^{\prime
}},\mathbf k-\mathbf k^{^{\prime }}\right) \right\}  \notag \\
& +i\sum_{k}\int dt\left\{ \eta ^{\ast }\left( t,\mathbf k\right) \mathfrak{z}\left(
t,\mathbf k\right) +\mathfrak{z}^{\ast }\left( t,\mathbf k\right) \eta \left( t,\mathbf k\right)
\right\} +F\left[ \eta ^{\ast },\eta \right] .  
\label{effact}
\end{align}
The equations of motion are obtained by minimizing the action with respect
to the variables $\mathfrak{z}$ and $\mathfrak{z}^{\ast }$, 
\begin{equation}
\left. \frac{\delta \left( S^{eff}\right) }{\delta \mathfrak{z}\left(
t,\mathbf p\right) }\right| _{\eta ^{\ast }=\eta =0}=0
\end{equation}
and
\begin{equation}
\eta\left( t,\mathbf x\right) =\eta_{x}\left( t,\mathbf x\right) +i\eta_{y}\left( t,\mathbf x\right)
\end{equation}
\ We find that 
the transverse components of the local 
magnetization satisfy the following
equations of motion 

\begin{align}
\frac{dS_{x}\left( t,\mathbf{p}\right) }{dt} & =\left( \Omega-K\right)
S_{y}\left( t,\mathbf{p}\right) -\eta_{y}\left( t,\mathbf{p}\right)  \notag
\\
& +\left(\frac{J}{2}\right)^{2}\sum_{\mathbf{k}}\int dt^{^{\prime}}\mathcal{K}\left( t,\mathbf{k}%
;t^{^{\prime}},\mathbf{k}+\mathbf{p}\right) \cos\left[ \left(
\varepsilon_{1}\left( \mathbf{k}\right) -\varepsilon_{2}\left( \mathbf{k+p}%
\right) \right) \left( t-t^{^{\prime}}\right) \right] S_{x}\left(
t^{^{\prime}},-\mathbf{p}\right)  \notag \\
& +\left(\frac{J}{2}\right)^{2}\sum_{\mathbf{k}}\int dt^{^{\prime}}\mathcal{K}\left( t,\mathbf{k}%
;t^{^{\prime}},\mathbf{k}+\mathbf{p}\right) \sin\left[ \left(
\varepsilon_{1}\left( \mathbf{k}\right) -\varepsilon_{2}\left( \mathbf{k+p}%
\right) \right) \left( t-t^{^{\prime}}\right) \right] S_{y}\left(
t^{^{\prime}},-\mathbf{p}\right),
\label{SxyEqs1}
\end{align}
and
\begin{align}
\frac{dS_{y}\left( t,\mathbf{p}\right) }{dt} & =-\left( \Omega+K\right)
S_{y}\left( t,\mathbf{p}\right) +\eta_{x}\left( t,\mathbf{p}\right)  \notag
\\
& -\left(\frac{J}{2}\right)^{2}\sum_{\mathbf{k}}\int dt^{^{\prime}}\mathcal{K}\left( t,\mathbf{k}%
;t^{^{\prime}},\mathbf{k}+\mathbf{p}\right) \sin\left[ \left(
\varepsilon_{1}\left( \mathbf{k}\right) -\varepsilon_{2}\left( \mathbf{k+p}%
\right) \right) \left( t-t^{^{\prime}}\right) \right] S_{x}\left(
t^{^{\prime}},-\mathbf{p}\right)  \notag \\
& +\left(\frac{J}{2}\right)^{2}\sum_{\mathbf{k}}\int dt^{^{\prime}}\mathcal{K}\left( t,\mathbf{k}%
;t^{^{\prime}},\mathbf{k}+\mathbf{p}\right) \cos\left[ \left(
\varepsilon_{1}\left( \mathbf{k}\right) -\varepsilon_{2}\left( \mathbf{k+p}%
\right) \right) \left( t-t^{^{\prime}}\right) \right] S_{y}\left(
t^{^{\prime}},-\mathbf{p}\right),
\label{SxyEqs2}
\end{align}
where 
\begin{equation}
\mathcal{K}\left( t,\mathbf{k};t^{^{\prime}},\mathbf{k}+\mathbf{p}\right)
=\Theta\left( t-t^{^{\prime}}\right) \left[ f\left( \varepsilon_{1}\left( 
\mathbf{k}\right) \right) -f\left( \varepsilon_{2}\left( \mathbf{k+p}\right)
\right) \right]
\end{equation}
and the correlation function for the random field  due 
to the conduction electrons is
given by 
\begin{align}
\left\langle \eta^{\ast}\left( t,\mathbf{p}\right) \eta\left(
t^{^{\prime}},\mathbf{p}^{^{\prime}}\right) \right\rangle & =\frac{J^{2}}{2}\left(
f\left( \varepsilon_{1}\left( \mathbf{p}\right) \right) -f\left(
\varepsilon_{2}\left( \mathbf{p}^{^{\prime}}\right) \right) \right) \coth%
\left[ \frac{\beta}{2}\left( \varepsilon_{2}\left( \mathbf{p}%
^{^{\prime}}\right) -\varepsilon_{1}\left( \mathbf{p}\right) \right) \right]
\\
& \times\exp\left[ -i\left( \varepsilon_{1}\left( \mathbf{p}\right)
-\varepsilon_{2}\left( \mathbf{p}^{^{\prime}}\right) \right) \left(
t-t^{^{\prime}}\right) \right].  \notag
\end{align}
\ The correlation function of the x-component is then given by 
\begin{equation}
{Re}\left\langle \eta_{x}\left( t,-\mathbf k\right) \eta_{x}\left(
t^{^{\prime}},\mathbf k^{^{\prime}}\right) \right\rangle =\left(\frac{J}{2}\right)^{2}\left( f\left(
\varepsilon_{1}\left(\mathbf  k\right) \right) -f\left( \varepsilon_{2}\left(\mathbf 
k^{^{\prime}}\right) \right) \right) \coth\left[ \frac{\beta
\omega_{k,k^{^{\prime}}}}{2}\right] \cos\left[ \omega_{k,k^{^{\prime}}}%
\left( t-t^{^{\prime}}\right) \right],
\end{equation}
where $\omega_{k,k^\prime}=\varepsilon_{1}\left(\mathbf  k\right) -
\varepsilon_{2}\left(\mathbf  k^\prime \right)$. 
\ This is one of the important results in this work. \ The kernel terms account 
for the dissipation and a shift in the frequency
due to the interaction with the conduction electrons.\ These 
equations of motion differ from the usual Gilbert form since 
 the dissipation term is not a derivative  
and is  non-local. \ These equations generalize
 those derived by Mills  for a Stoner 
particle at zero temperature and in the adiabatic limit.\cite{mills} \ Hence 
the memory terms will be very important for local moments.
\ In the next section 
we study the limit under which the GB equation is recovered in a thin film embedded between 
two large reservoirs at equilibrium. \

\section{Finite-size effects in the memoryless limit}

\begin{figure}[ht]
  \begin{center}
 \mbox{\epsfig{file=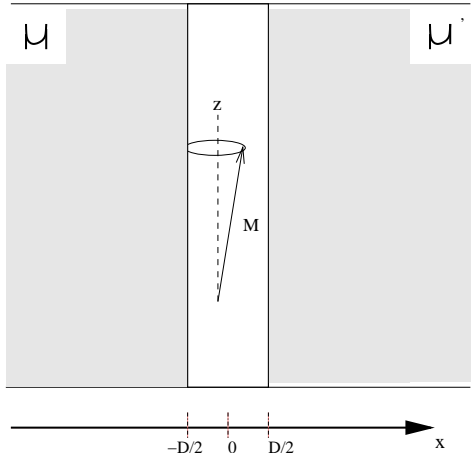,height=6 cm}}
  \end{center}
  \caption{{{A thin film confined between two large reservoirs with 
the same chemical potentials $\mu = \mu^\prime$.\ The case 
of different $\mu$'s is treated in sect. IV. }}}
 \label{reservoir}
\end{figure}

In a series of very illuminating papers, Simanek was able to show how 
the ideas of Tserkovniak, Brataas and Bauer can be understood in the 
familiar linear response approach which avoids the use of the 
scattering method. \cite{simanek, simanek2,tser}  \ Similar 
calculations were carried out by Mills \cite{mills}
using a dynamic RKKY approach which generalizes
 the earlier results obtained by Berger 
\cite{berger}. \ It is well known that the spin-pumping theory and
 the Berger-Mills theory both give interfacial additional damping due
to spin currents (not charge currents). \  In the former 
theory, this damping vanishes 
when there is no exchange splitting between the spin-up and spin-down 
electrons while it does not in the Berger-Mills theory. \ The spin 
pumping theory however seems to be very successful in interpreting 
the recent experiments by Mizukami et al. \cite{mizukami}. \ Hence in this 
section, we use the equations derived in the previous section
 to further understand 
this particular discrepancy between the various 
methods. \  Our results 
happen to be similar to those derived by Simanek using the 
spin-pumping 
theory. \ We believe however that our approach is more direct and 
transparent 
besides it is self-contained. \ This equivalence has very important 
consequences on the understanding of the  physical 
origin of the spin momentum torque in finite films. \ Moreover, our 
theory is easy to extend to finite temperature and can deal 
with transient conditions as we show in the next section.

\ First, let's set-up the geometry of the problem and 
calculate the damping in the limit 
 when there is no memory in the magnetic 
system, i.e., the average
magnetization is much slower than the conduction electrons. \ This is 
the adiabatic limit. \ The geometry we adopt (fig. \ref{reservoir} 
is the same as the one adopted 
by TBB \cite{tser}. \ The two reservoirs on each side of the thin film
will act as a sink for the spin leaked through the interfaces. \ The 
reservoirs are maintained at the same chemical potential $\mu$ in 
this section and 
hence there is no net flow of charge from left to right. \ Our theory 
can be also adapted to the case of non-equal chemical potentials which 
is briefly addressed in the following section.

For the rest of this section we use Mills \cite{mills} 
notation since he was able to derive a more general 
form for the Gilbert equation in the adiabatic limit. \ We will 
show below how our theory reduces to his in this limit.

 \ Within the linear response approach, the Gilbert equation for
the magnetization $\mathbf{M}$ is 
\begin{eqnarray}
\frac{\partial \mathbf{M}}{\partial t} &=&-\left| \gamma \right| \left[ 
\mathbf{M}\times \left( \mathbf{H}+\left\langle \mathbf{H}%
_{eff}(t)\right\rangle \right) \right]   \notag \\
&&+\frac{G\left| \gamma \right| }{\gamma ^{2}M_{s}^{2}}\left[ \mathbf{M}%
\times \frac{\partial \mathbf{M}}{\partial t}\right] ,
\end{eqnarray}
where $G$ is the Gilbert constant and the effective field $\mathbf{H}_{eff}$
is due to the action of the conduction electrons with the magnetization 
through a sd-type interaction 
\begin{eqnarray}
H_{x}^{eff}(t) &=&\frac{J^{2}V_{c}}{2N\hbar \mu _{0}^{2}}\left( \wedge _{2}%
\frac{dM_{x}}{dt}+\wedge _{1}\frac{dM_{y}}{dt}\right) , \\
H_{y}^{eff}(t) &=&\frac{J^{2}V_{c}}{2N\hbar \mu _{0}^{2}}\left( \wedge _{2}%
\frac{dM_{y}}{dt}-\wedge _{1}\frac{dM_{x}}{dt}\right) .  \notag
\end{eqnarray}
Hence the conduction electrons enhance the Gilbert 
damping term by
\begin{equation}
\Delta G=\frac{J^{2}V_{c}}{2N\hbar \mu _{0}^{2}}\wedge _{2}\left| \gamma
\right| ^{2}M_{s}^{2}.
\end{equation}
It was moreover argued by Mills that the constant $\wedge _{1}$, which
renormalizes the precessional frequency, is not zero in general as was
assumed by Simanek and Heinrich \cite{simanek}. \ Below we show
that our analysis naturally gives an {\it explicit} expression for 
this term and
 that it vanishes within the approximations  
employed here. \ In higher orders in $J^2$, the contribution 
of this term is non-zero but small as we show in the next section.

From the equations of motion for the x,y-components, 
Eqs. \ref{SxyEqs1}-\ref{SxyEqs2}, we 
easily see that it is the term that has the sine-dependence 
that gives rise to 
damping, 

\begin{eqnarray}
\Delta G_{x(y)}\left( t,\mathbf p\right)  &=&\frac{J^{2}}{2}\int_{0}^{\infty }dt^{\prime
}\sum_{k}\left[ f(\varepsilon _{1}(\mathbf k)-f\left( \varepsilon _{2}(\mathbf k + \mathbf p)\right)
\right]   \label{dgxy} \\
&&\times \sin \left[ \left( \varepsilon _{1}(\mathbf k)-\varepsilon _{2}(\mathbf k + \mathbf p)\right)
t^{^{\prime }}\right] S_{y(x)}\left( t-t^{^{\prime }},-\mathbf p\right) .  \notag
\end{eqnarray}
The cosine-dependent term will be shown below to be the origin of 
the term $\wedge_{1}$ discussed by Mills. \
If we use the slow (adiabatic) precession approximation for the spin variables,
\begin{equation}
S_{x(y)}\left( t-t^{^{\prime }},-\mathbf p\right) =S_{x(y)}\left( t,-\mathbf p\right)
-t^{\prime }\frac{dS_{x(y)}\left( t,-\mathbf p\right) }{dt}+...
\end{equation}
Inserting this back in Eq. \ref{dgxy}, we get 
\begin{equation}
\Delta G_{x(y)}\left( t,\mathbf p\right) =\Delta G_{x(y)}^{(p)}\left( t,\mathbf p\right)
+\Delta G_{x(y)}^{(d)}\left( t,\mathbf p\right) ,
\end{equation}
with
\begin{eqnarray}
\Delta G_{x(y)}^{(p)}\left( t,\mathbf p\right)  &=&\frac{J^{2}}{2}S_{y(x)}\left( t,-\mathbf p\right)
\sum_{\mathbf{k}}\int_{0}^{\infty }dt^{\prime }\left[ f(\varepsilon
_{1}(\mathbf k )-f\left( \varepsilon _{2}(\mathbf k + \mathbf p)\right) \right]   \notag \\
&&\times \sin \left[ \left( \varepsilon _{1}(\mathbf k)-\varepsilon _{2}(\mathbf k + \mathbf p)\right)
t^{^{\prime }}\right] 
\end{eqnarray}
and
\begin{eqnarray}
\Delta G_{x(y)}^{(d)}\left( t,\mathbf p\right)  &=&-\frac{J^{2}}{2}\frac{dS_{y(x)}\left(
t,-\mathbf p\right) }{dt}\sum_{\mathbf{k}}\int_{0}^{\infty }dt^{\prime }\left[
f(\varepsilon _{1}(\mathbf k)-f\left( \varepsilon _{2}(\mathbf k + \mathbf p)\right) \right]   \notag
\\
&&\times t^{\prime }\sin \left[ \left( \varepsilon _{1}(\mathbf k)-\varepsilon
_{2}(\mathbf k + \mathbf p)\right) t^{^{\prime }}\right] .
\end{eqnarray}
The term $\Delta G_{x(y)}^{(p)}$ contributes to the precessional frequency
while \ $\Delta G_{x(y)}^{(d)}$ gives rise to Gilbert type damping.\ Hence
it should be related to the imaginary part of a susceptibility term. \ In
fact we can write 
\begin{equation}
\Delta G_{x(y)}^{(d)}\left( t,\mathbf p\right) =\left. -i\frac{d\chi \left( \Omega
,\mathbf p\right) }{d\Omega }\right| _{\Omega =0}\frac{dS_{y(x)}\left( t,-\mathbf p\right) }{%
dt},
\end{equation}
where the 'susceptibility' function $\chi $ is
\begin{equation}
\chi \left( t,\mathbf p\right) =-\frac{J^{2}}{2}\Theta \left( t\right) \sum_{\mathbf{k}}\left[
f(\varepsilon _{1}(\mathbf k))-f(\varepsilon _{2}(\mathbf k + \mathbf p))\right] \sin \left[ t\left(
\varepsilon _{1}(\mathbf k)-\varepsilon _{2}(\mathbf k + \mathbf p)\right) \right] ,
\end{equation}
and its Fourier transform  
\begin{eqnarray}
\chi \left( \Omega ,\mathbf p\right)  &=&\int_{-\infty }^{\infty }\chi \left(
t,\mathbf p \right) e^{i\Omega t}dt  \notag \\
&=&-\frac{J^{2}}{4}\sum_{\mathbf{k}}\left[ f(\varepsilon _{1}(\mathbf k)-f\left( \varepsilon
_{2}(\mathbf k + \mathbf p)\right) \right]   \notag \\
&&\times \left[ \frac{1}{\varepsilon _{1}(\mathbf k)-\varepsilon _{2}(\mathbf k+\mathbf p)+\Omega
+i\eta }+\frac{1}{\varepsilon _{1}(\mathbf k)-\varepsilon _{2}(\mathbf k+\mathbf p)-\Omega -i\eta }
\right] ,
\end{eqnarray}
where $\eta $ is a small positive real number. From this expression, we see
that to allow for a finite relaxation time $\tau _{s}$ of the conduction
electrons, we can just replace $\Omega $ by $\Omega +i/\tau _{s}$. The
constants $\Lambda _{2}$ introduced by Mills can be 
obtained from 
$\chi $ through the following expression
\begin{equation}
\left. -i\frac{d\chi \left( \Omega ,\mathbf p\right) }{d\Omega }\right| _{\Omega
=0}=\Lambda _{2r}(\mathbf p)+i\Lambda _{2}(\mathbf p).
\end{equation}
An easy calculation shows that $\Lambda _{2r}$ vanishes within this
approximation and $\Lambda _{2}$, which is still a function of the momentum,
is given by
\begin{eqnarray}
\Lambda _{2}\left( \mathbf p\right)  &=&-J^{2}\sum_{\mathbf{k}}\left[ f(\varepsilon
_{1}(\mathbf k))-f(\varepsilon _{2}(\mathbf k+\mathbf p))\right] \left( \varepsilon
_{1}(\mathbf k)-\varepsilon _{2}(\mathbf k+\mathbf p)\right)   \notag \\
&&\times \left[ \frac{\eta }{\left( \left( \varepsilon _{1}(\mathbf k)-\varepsilon
_{2}(\mathbf k+\mathbf p)\right) ^{2}+\eta ^{2}\right) ^{2}}\right] .
\end{eqnarray}
As shown by Heinrich, Fraitova and Kambersky \cite{heinrich} in an infinite
medium, the conduction electrons can't dissipate energy unless the 
electron-hole pairs
have a finite-lifetime by transferring
 energy to the lattice. \ This can be 
done by taking a temperature-dependent 
finite $\eta$. \ In this case the damping 
term $\Lambda _{2}\left( \mathbf p\right) $ will not vanish. \ This spin-flipping 
mechanism is believed to be the source of the damping in iron and 
permalloy 
\cite{bahagat}. \ Another source for damping can be geometrical in 
origin. \ As shown by Mills a breakdown of wave vector conservation 
due of the finite size of the film can give rise to dissipation. \ In the 
language of reservoirs, we rephrase this by saying that quenching the 
states of the magnetic film to a countable number while leaving those 
of the electronic states denumerable is equivalent to a Caldeira-Leggett (CL) 
model with a fermionic bath 
which is known to give rise to quantum dissipation.\cite{CL,rebei} \ Our 
geometry is then a typical example of this model and should 
show a dissipative behavior as a function of the thickness $D$, i.e., 
 $\Lambda
_{2}\left( p\right) \rightarrow 0$ as $D \rightarrow \infty$, even in the case of
 very slow relaxation times for the conduction electrons.  \ We take the 
thin film
to have finite
 thickness $D$ in the x-direction, and we take  the transverse
components, $S_{x}$ and $S_{y}$ to dependent only on the x-coordinate normal
to the plane with pinned or unpinned
 boundary conditions. \ In the continuum approximation, the 
magnetization 
components in this symmetric configuration take the form
\begin{equation}
 S_{i}\left( t, \mathbf{r}\right) = \sum_{n=0,1,2,...} S_{i}^{n}(t) 
\; \cos\left( \frac{n \pi}{D} x \right), \; \; \; i =x,y
\end{equation}
The Fourier transform is given
by 
\begin{equation}
\int dx \; S_{i}\left( t,x\right) e^{ipx}=S_{i}\left( t,p\right) 
\end{equation}
therefore, we have $S_{i}\left( t,-p\right) =S_{i}^{\ast }\left( t,p\right) $
and  
\begin{equation}
S_{i}\left(  t,\mathbf{p}\right)  =\left(  2\pi\right)  ^{2}\delta\left(
\mathbf{p}_{||}\right)  \sum_{n}S_{i}^{n}\left(  t\right)  \left[  \frac
{\sin\left(  \left(  p+\frac{n\pi}{D}\right)  D/2\right)  }{\left(
p+\frac{n\pi}{D}\right)  }+\frac{\sin\left(  \left(  p-\frac{n\pi}{D}\right)
D/2\right)  }{\left(  p-\frac{n\pi}{D}\right)  }\right]  ,
\end{equation}
where we have set $p_{x}=p,-\infty<p<\infty$.

\ In the following we assume that the splitting $\delta$ in electronic energy
bands
due to the sd-interaction is  smaller than the Fermi energy and
 $k_BT << \mu$ which is the case at room 
temperature, then it is enough for our purposes (because of the finite size)
 to use the following  
 approximation for the Fermi-Dirac functions for the conduction electrons 
\begin{eqnarray}
\left[ f\left( \varepsilon _{2}(\mathbf k+\mathbf p)\right) -f(\varepsilon _{1}(\mathbf k)\right]
&\cong &\frac{\partial f}{\partial \varepsilon _{\mathbf k}}\left( \varepsilon _{2}(%
\mathbf{k}+\mathbf{p})-\varepsilon _{1}(\mathbf{k})\right)   \notag \\
&=&\frac{\partial f}{\partial \varepsilon _{\mathbf{k}}}|_{
\overline\varepsilon_F}  \left( 
\frac{\mathbf{k}\cdot \mathbf{p}}{m}+\delta \right) ,
\label{f2f1}
\end{eqnarray}
where $\overline{\varepsilon}_F = \left( \varepsilon_F^{\uparrow} + 
\varepsilon_F^{\downarrow} \right)/2$. \ This approximation is not 
necessary and will not change our 
conclusions but it helps keep the algebra at minimum, otherwise 
 the Lindhard function will appear explicitly in our expressions 
and will considerably add to the complexity of the calculations. \ If 
 a non-zero voltage difference is applied across 
the thin film, then 
 Eq. \ref{f2f1} has to be modified to take account of
the spin accumulation effect 
 $\Delta^{\uparrow\downarrow}=\mu^\uparrow - \mu^\downarrow$ 
 due to the normal-ferromagnetic interface 
\cite{vanson}. \ This term which can be positive or negative depending on the direction 
of the polarized current will hence contribute to the damping. \ We will 
say more about this case when we study the noise and the 
corresponding fluctuation dissipation theorem in the next 
section. \ Using the approximation Eq. \ref{f2f1}, 
the damping term becomes   (for $p \neq 0$)
\begin{eqnarray}
\wedge _{2}\left( p\right)  &=&\frac{ J^{2} v}{2\left( 2\pi \right) ^{2}\tau
_{s}}\frac{m}{p\varepsilon _{F}}\left[ \frac{\delta -p\sqrt{\frac{%
2\varepsilon _{F}}{m}}}{\left[ \left( \delta -p\sqrt{\frac{2\varepsilon _{F}%
}{m}}\right) ^{2}+\tau _{s}^{-2}\right] }-\frac{\delta +p\sqrt{\frac{%
2\varepsilon _{F}}{m}}}{\left[ \left( \delta +p\sqrt{\frac{2\varepsilon _{F}%
}{m}}\right) ^{2}+\tau _{s}^{-2}\right] }\right.   \notag \\
&&\left. -\tau _{s}\left( \tan ^{-1}\left[ \tau _{s}\left( \delta -p\sqrt{%
\frac{2\varepsilon _{F}}{m}}\right) \right] -\tan ^{-1}\left[ \tau
_{s}\left( \delta +p\sqrt{\frac{2\varepsilon _{F}}{m}}\right) \right]
\right) \right] ,
\end{eqnarray}
where the Fermi energy $\varepsilon _{F}$ is that of the spin up electron in
the ferromagnet. \ Care is needed to get the
 corresponding expression for $p=0$.\ The damping is therefore momentum-dependent 
as we should expect in a 
finite film. \ This expression can be, e.g., useful 
for studies of spin-wave resonance in thin films. \ In the remaining, we 
confine ourselves to the volume mode since it is usually 
the mode measured by FMR. \ In this case, the damping simply
becomes
\begin{equation}
\overline{\wedge }_{2}\simeq \frac{4}{\pi D}\int_{0
}^{\infty}dp \, \left( \frac{\sin \left( p\frac{D}{2}\right) }{p}\right)
^{2}\wedge _{2}(p).
\end{equation}
As it is clear from this expression, damping is directly 
related to the breakdown of momentum conservation in the direction 
normal to the film. \ This damping clearly vanishes when the size
of the film becomes infinite, i.e., 
 $D \rightarrow \infty$. \ This expression for the damping 
was found by  Fourier transforming back the equations of motion 
, eqs. \ref{SxyEqs1}-\ref{SxyEqs2},
to real-space and enforce the condition that 
the magnetization vanishes for $|x| > D/2$. \ This is the 
procedure that we followed to  allow
 us to capture the finite size effects
of the magnetic film on the damping in this section and on 
the noise 
 in the next one. \ Using the fact 
that the saturation 
magnetization is defined by $M_s = \mu_0/v$, the excess Gilbert 
damping is therefore given by
\begin{eqnarray}
 G_s = \frac{\mu_0^2}{v}\overline{\wedge }_{2}.
\end{eqnarray}
We write this in terms of small dimensionless 
parameters $\delta _{r}=\frac{%
\delta }{\varepsilon _{F}}$ and $\tau _{r}=\frac{1}{\tau _{s}}/\varepsilon
_{F}$ with typical values 0.4 and 0.001, respectively in transition metals. 
The damping is now given by
\begin{equation}
G_{s}=\delta_{r}^{2}\frac{\mu_{0}^{2}k_{F}^{2}}{2\pi^{2}\hbar D}F\left(
\tau_{r},\delta_{r},D\right),
\end{equation}
with
\begin{align}
F\left(  \tau_{r},\delta_{r},D\right)    & =\frac{\tau_{r}}{\pi}\int
_{0}^{\infty}dx\frac{1}{x^{2}}\sin^{2}\left(  \frac{1}{4}k_{F}Dx\right)
\left\{  \frac{\delta_{r}-x}{x\left[  \left(  \delta_{r}-x\right)  ^{2}%
+\tau_{r}^{2}\right]  }\right.  \nonumber\\
& -\frac{\delta_{r}+x}{x\left[  \left(  \delta_{r}+x\right)  ^{2}+\tau_{r}%
^{2}\right]  }-\frac{\tan^{-1}\left[  \left(  \delta_{r}-x\right)  /\tau
_{r}\right]  }{x\tau_{r}}\nonumber\\
& \left.  +\frac{\tan^{-1}\left[  \left(  \delta_{r}+x\right)  /\tau
_{r}\right]  }{x\tau_{r}}\right\}
\end{align}
In the limit of vanishing bulk damping, $\tau_{r}\rightarrow0 $, 
and large thickness $D$, the damping is given by
\begin{equation}
G_{s}=\frac{\mu_{0}^{2}k_{F}^{2}}{2\pi^{2}\hbar}\left(  \frac{0.55\delta_{r}%
}{D}\right).
\end{equation}
To get this result, we used the approximation
\begin{equation}
f ( x ) = \int_{x}^{\infty} dy \frac{sin^2 y}{y^2} \approx \frac{0.55}{x},
\end{equation}
for large $x$. \ Hence this damping 
is due to finite-size effects. \ At smaller thicknesses, the 
damping shows some oscillatory behavior as shown in
 figs \ref{damp1},\ref{damp2},\ref{damp3} for small bulk 
damping. \  The size ($\approx 10^8 \; sec^{-1}$) and the oscillatory
 behavior of the damping  near the interface
are similar to what
 Mills \cite{mills} and Simanek \cite{simanek3}
 found and is non-existent in a large 
sample as is clear from the figure for large $D$. \ The dimensionless 
Gilbert damping for iron would be therefore 
\begin{equation}
\alpha = \frac{G_s}{\gamma M_s} \approx 0.01,
\end{equation}
which has the right order of magnitude as measured in ref. 
\onlinecite{mizukami}. \ Figure \ref{damp5} is perhaps the case that applies
to transition metals. \ In this case the oscillations are almost 
nonexistent and is consistent with the recent numerical
calculations of Zwierzycki et al. \cite{zw} which are based on the circuit 
theory approach. \cite{brataas} \ The temperature dependence 
is weak in all the results since we have 
assumed that $k_BT << \epsilon_F$. \ For large thicknesses, the damping is therefore still dependent on the exchange coupling 
and this dependence is linear. \ This is in contrast to Berger's \cite{berger}
 result where his interfacial
damping is independent of $J$.

\begin{figure}[ht]
\begin{center}
 \mbox{\epsfig{file=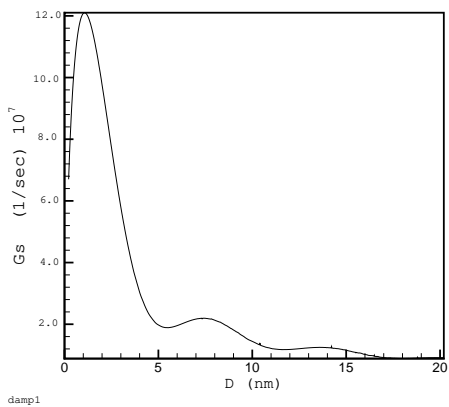,height=6 cm}}
\end{center}
\caption{{The damping constant (in $sec^{-1}$) as a function of 
the thickness of the ferromagnetic film (in $nm$). $k_F = 10^8$ cm$^{-1}$,
 $\delta_r=0.2$, $\eta_r=0.001$}}
\label{damp1}
\end{figure}

\begin{figure}[ht]
\begin{center}
 \mbox{\epsfig{file=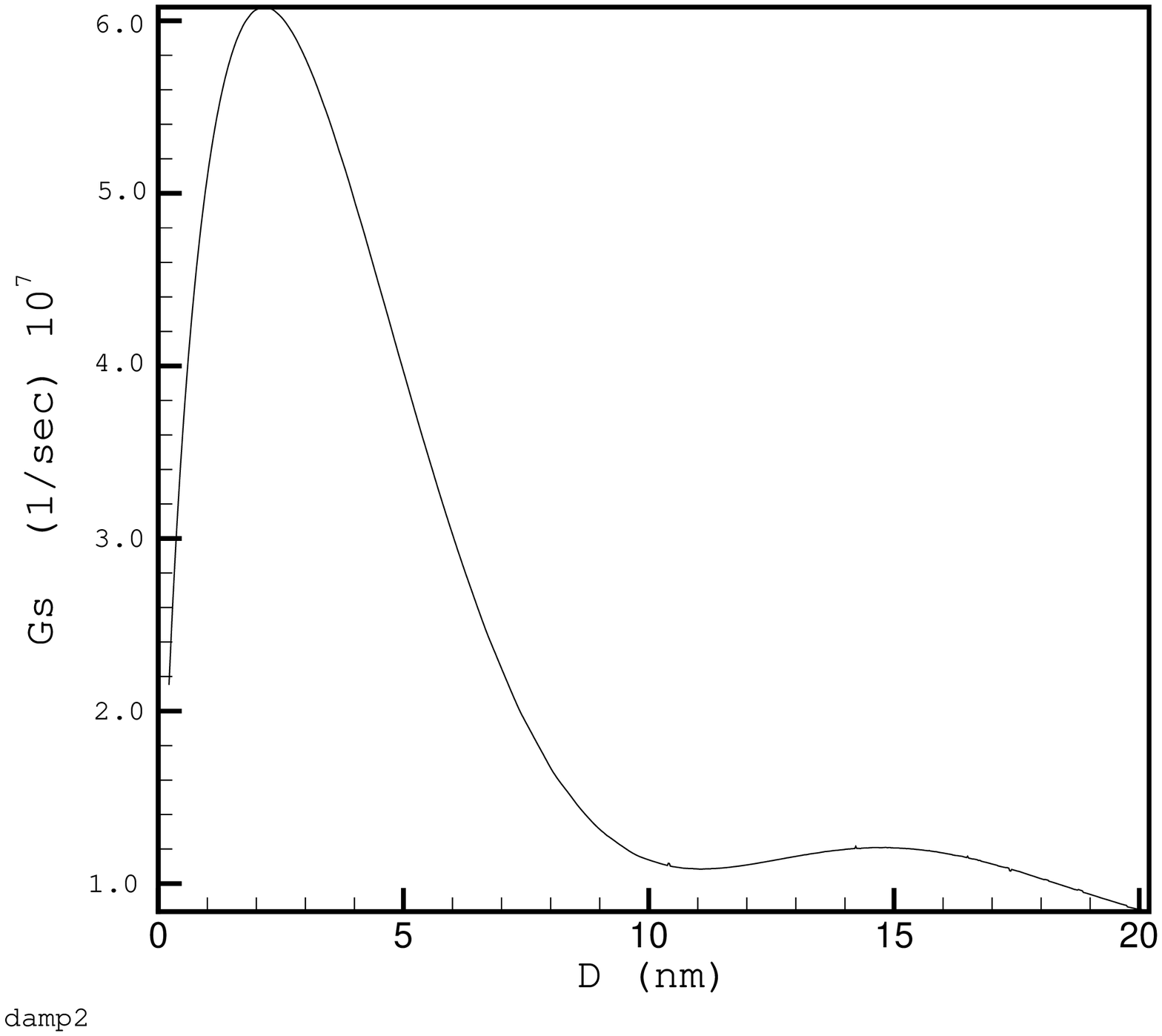,height=6 cm}}
\end{center}
\caption{{The damping constant (in $sec^{-1}$) as a function of 
the thickness of the ferromagnetic film (in $nm$). $k_F = 10^8$ cm$^{-1}$,
 $\delta_r=0.1$, $\eta_r=0.001$.}}
\label{damp2}
\end{figure}

\begin{figure}[ht]
\begin{center}
 \mbox{\epsfig{file=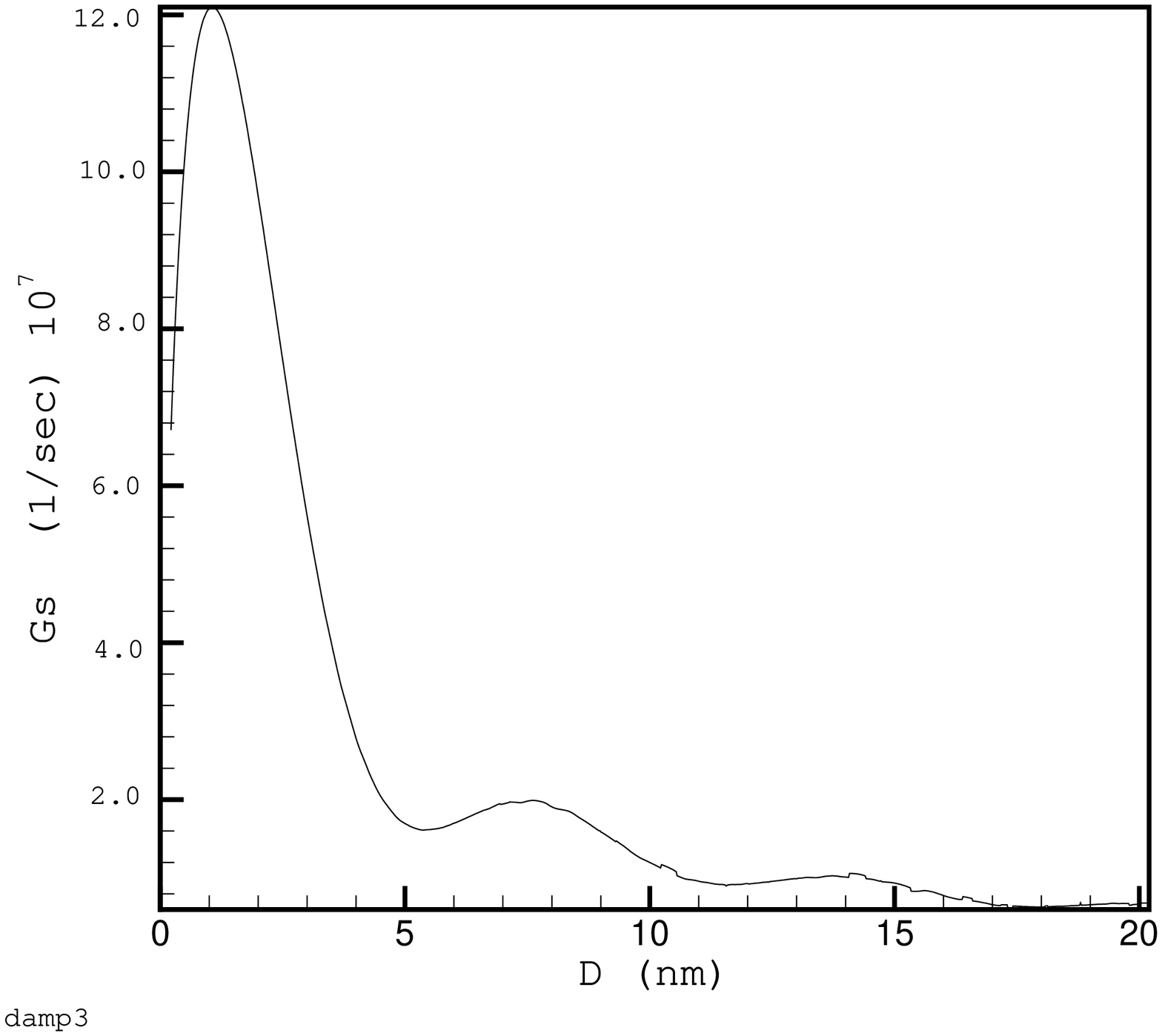,height=6 cm}}
\end{center}
\caption{{The damping constant (in $sec^{-1}$) as a function of 
the thickness of the ferromagnetic film (in $nm$). $k_F = 10^8$ cm$^{-1}$,
 $\delta_r=0.2$, $\eta_r=0.001$}}
\label{damp3}
\end{figure}

\begin{figure}[ht]
\begin{center}
 \mbox{\epsfig{file=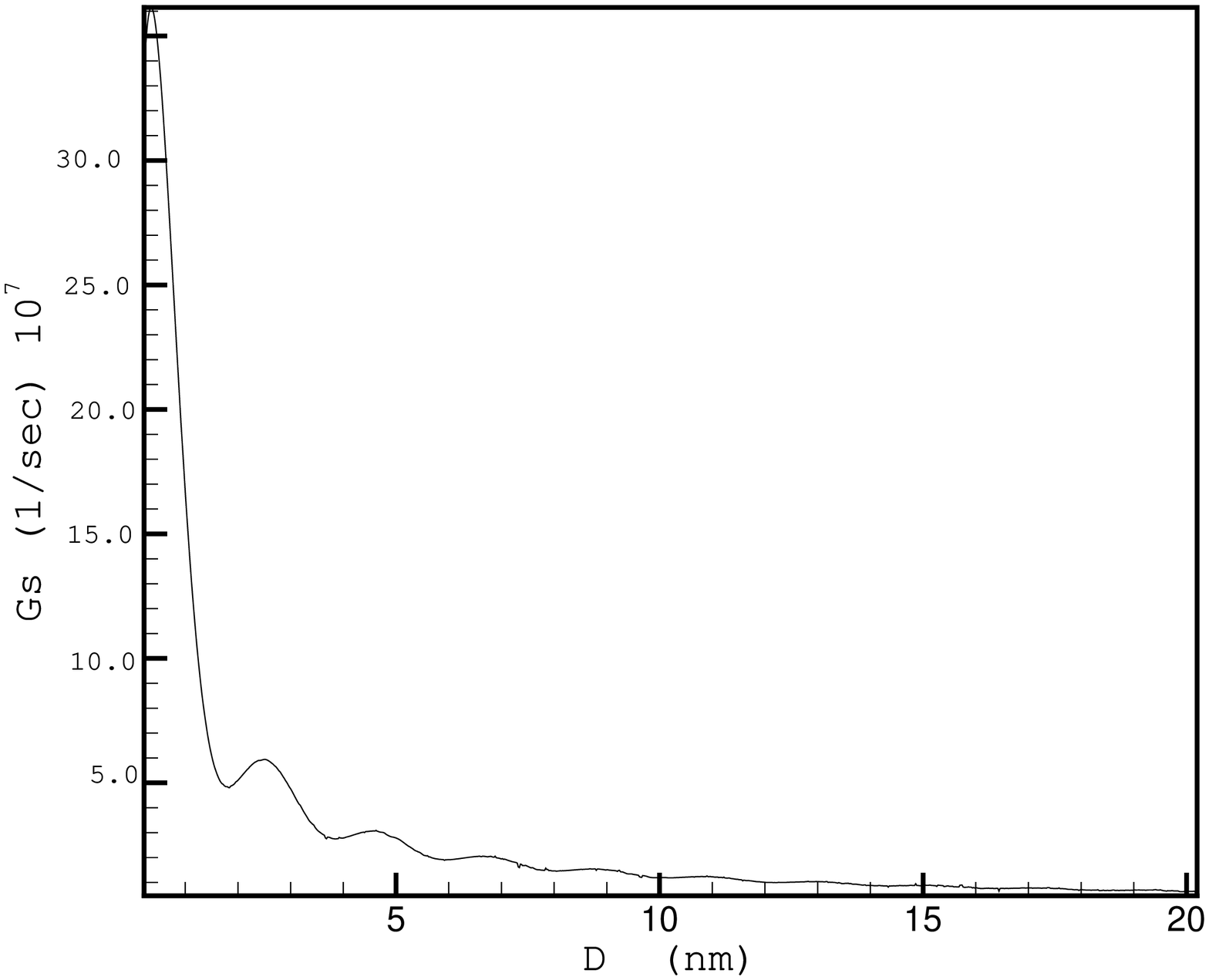,height=6 cm}}
\end{center}
\caption{{The damping constant (in $sec^{-1}$) as a function of 
the thickness of the ferromagnetic film (in $nm$). $k_F = 10^8$ cm$^{-1}$,
 $\delta_r=0.6$, $\eta_r=0.001$}}
\label{damp5}
\end{figure}

\ Finally, we show that the term $\wedge_1$ vanishes to 
lowest order. \ 
To find an explicit expression for the $\ \Lambda_{1}$-term, we
proceed along similar lines as we did for $\Lambda_{2}$. We make a slow time
approximation 
for the cosine-term in eq. \ref{SxyEqs1} and we define a 'susceptibility'
function$\ \ \chi^{\prime}\left(  \Omega,\mathbf{p}\right)  $ similar to
$\chi$,%
\begin{eqnarray}
\chi^{\prime}\left(  \Omega,\mathbf{p}\right)  & = & - i J^{2}\sum_{\mathbf{k}%
}\left(  f\left(  \varepsilon_{1}\left(  \mathbf{k}\right)  \right)  -f\left(
\varepsilon_{2}\left(  \mathbf{k}+\mathbf{p}\right)  \right)  \right)  \\
&& \times\left[
\frac{1}{\varepsilon_{1}\left(  \mathbf{k}\right)  -\varepsilon_{2}\left(
\mathbf{k}+\mathbf{p}\right)  +\Omega+i\eta}-\frac{1}{\varepsilon_{1}\left(
\mathbf{k}\right)  -\varepsilon_{2}\left(  \mathbf{k}+\mathbf{p}\right)
-i\eta}\right]  . \nonumber
\end{eqnarray}
As in the $\Lambda_{1}$ case, $\Lambda_{2}$ is proportional to the imaginary
part of $d\chi^{\prime}\left(  \Omega=0,\mathbf{p}\right)  /d\Omega$ which is
easily seen to vanish. \ At higher orders 
in $J^{2}$ it gives a nonzero
contribution to the frequency. \ This is the subject of next section.

\section{the magnetic noise spectrum}

\begin{figure}[ht]
\begin{center}
 \mbox{\epsfig{file=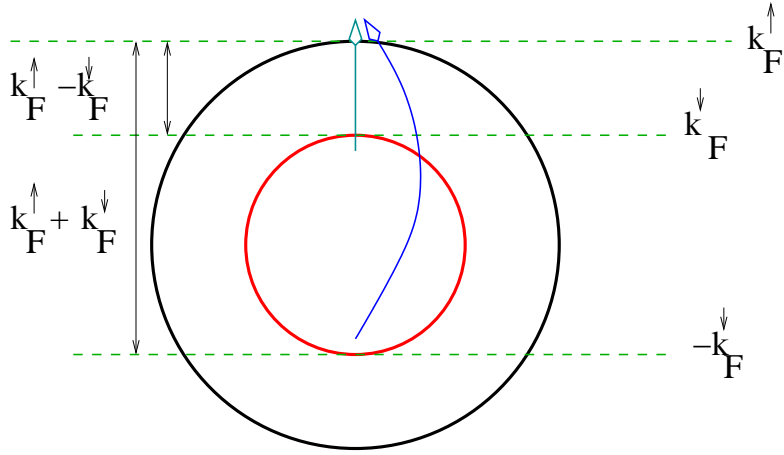,height=6 cm}}
\end{center}
\caption{A spin-down/spin-up excitation in a bulk ferromagnet with smallest 
and largest wave-vectors.}
\label{fermi}
\end{figure}

In this section, we calculate the various correlation 
functions 
 of the 
magnetization vector by including higher
 order corrections in $J$ in the 
exchange field. \ We will deal with
both the low frequency limit and the 
high frequency regime. \ The former is applicable to the 
case of large magnetization while the latter is important 
for local atomic moments. \ We will show that the damping 
for local moments depends on the symmetry of the Hamiltonian 
and will not be of the Gilbert-form. \ In the adiabatic limit
which applies to the average magnetization of the film, the GB 
equation is recovered. 

\ This is a direct extension of the calculations 
presented in the previous sections. \ In the bulk it was 
shown  by Heinrich, Fraitova and Kambersky 
\cite{heinrich} that the sd-exchange gives zero damping 
unless dissipation to the phonons is included. \ This was done
in a non-self-consistent way 
 by putting  a relaxation time $\tau_s$ by 
hand in the electron propagator. \ However, based on a simple 
analogy with Migdal's theory \cite{migdal}
 on electrons and phonons, we should
expect a nonzero damping for spin waves with wave numbers $p$ such 
that $0 < k_F^{\uparrow}-k_F^{\downarrow} < p < 
k_F^{\uparrow}+k_F^{\downarrow}$ if $k_B T << \epsilon_F$ 
(fig. \ref{fermi}). \ Hence 
the volume mode, i.e, $ p = 0$ mode, won't be expected 
to show any dissipation in the bulk. \ Sine we are partly
 interested 
in long-wavelength excitations in thin films, we 
will concentrate on the
consequences of
interfacial effects which based on the calculations above 
should provide a new source for dissipation.  \ We follow 
closely Schwinger's original 
work on the harmonic 
oscillator interacting with a bath of harmonic 
oscillators.\cite{schwinger} \ Some of the computations
will be deferred to the appendix in the hope not to distract
the reader from the end results. \  Our treatment is self-contained. 
 \ Similar 
calculations have been carried out in Ref. \onlinecite{rebei} which 
dealt with a 
hypothetical physical model for dissipation in the 
bulk. \ However in the adiabatic limit, we will recover the results 
in \onlinecite{rebei}. \ This shows that 
the macroscopic magnetization as opposed to the local 
moments is insensitive to the dynamics of the 
environment.\ In this section, the bulk spin 
relaxation time $\tau_s = 0$. \ We go beyond the mean field 
approximation for the exchange field. \cite{rebei6} \ In other words, 
we seek corrections to the propagators of the theory by 
including self-energy corrections. \ This will allow us 
to go beyond the approximations made in the 
previous sections and calculate the damping due to 
 inelastic 
scattering of the conduction electrons off the 
magnons as they cross the interfaces. \ We  do  not  use 
the slow-time approximation  in this section, but the damping
will be shown to have the Gilbert form for frequencies much 
smaller than the 
electronic precessional frequency. \ Moreover, we find a very 
interesting result that relates the 
GB equation for spin-momentum transfer
to the model treated here. \ We show that the Langevin dynamic 
treatment
of 
Li
and Zhang \cite{li} for the noise 
can be only justified for the adiabatic limit, 
 and around  the FMR frequency in thin films. 
\ As in the previous section, the magnetization precesses around an 
effective in-plane field directed along the z-axis. \ We start by 
coupling the transverse
magnetization $\mathbf{S}=\left(  S_{x},S_{y}\right)$ to external 
sources  $\mathbf{J}_{1}$ and $\mathbf{J}_{2}$\ along the
positive and negative-time oriented paths, respectively. The modified
Hamiltonian is now given by 
\begin{equation}
\mathcal{H}_{P}=\mathcal{H}_{1}-\mathbf{J}_{1}\cdot\mathbf{S}_{1}-\left(
\mathcal{H}_{2}-\mathbf{J}_{2}\cdot\mathbf{S}_{2}\right).
\end{equation}
Next we  introduce the new variables $\mathbf{J}=\frac{1}{2}\left(  \mathbf{J}_{1}+\mathbf{J}_{2}\right)  $ and $\mathbf{Q}=\mathbf{J}_{1}-\mathbf{J}_{2} $. \
Similarly, we define an average $\mathbf{S}=\frac{1}{2}\left(  \mathbf{S}_{1}+\mathbf{S}_{2}\right)  $ and a difference $\mathbf{D}=\mathbf{S}_{1}-\mathbf{S}_{2} $
representing the fluctuations in the magnetization. \ 
The generating functional $\mathbb{Z}\left[  \mathbf{J}_{1},\mathbf{J}%
_{2}\right]  $ is defined in such way that its Taylor expansion at
$\mathbf{J}_{1}=\mathbf{J}_{2}=0$ gives the correlation functions of the
magnetization,
\begin{equation}
\mathbb{Z}\left[  \mathbf{J}_{1},\mathbf{J}_{2}\right]  =Tr\left(  \rho
T^{-1}\exp\left[  i\int_{t_{0}}^{t_{f}}dt\mathcal{H}\left[  \mathbf{J}%
_{1}\right]  \right]  T\exp\left[  -i\int_{t_{0}}^{t_{f}}dt\mathcal{H}\left[
\mathbf{J}_{2}\right]  \right]  \right)  .
\end{equation}
Then we see from the definitions that%
\begin{equation}
\left.  \frac{\delta\log\mathbb{Z}\left[  \mathbf{J}_{1},\mathbf{J}%
_{2}\right]  }{\delta\mathbf{J}_{1}(t)}\right|  _{J_{1}=J_{2}=0}=i\left\langle
\mathbf{S}_{1}(t)\right\rangle ,
\end{equation}
and
\begin{align}
\frac{\delta\left\langle S_{i}(t)\right\rangle }{\delta Q_{j}(t^{\prime})}  &
=i\left\langle S_{i}(t)S_{i}(t)\right\rangle \nonumber\\
& =\frac{i}{2}\left\langle \left\{  \widehat{S}_{i}(t),\widehat{S}%
_{j}(t^{\prime})\right\}  \right\rangle .
\end{align}
The latter symmetric average is the one usually associated with the 
noise in the magnetization vector. \ We will seek a general 
expression for this quantity that takes into account initial 
conditions, i.e., the reservoirs and the film are assumed initially 
to be separately in equilibrium before they are put in 
contact with each other at time $t_0$. \ Hence our method is capable of 
handling transient behavior in our system.
\bigskip To calculate the symmetric correlation function, we need first to
find the equation of motion of the average value $\left\langle \mathbf{S}%
\right\rangle $ . \ For this, we need the effective action in the presence of
the external fields  which is easily found from Eq. \ref{effact}%

\begin{align}
iS_{eff}\left[  \mathfrak{Z}^{\ast},\mathfrak{Z,z}^{\ast}\mathfrak{,z}\right]    &
=\sum_{\mathbf{k}}\int_{t_{0}}^{t_{f}}dt\left[  -\frac{1}{2}\mathfrak{Z}^{\ast
}\left(  \mathbf{k}\right)  \partial_{t}\mathfrak{z}\left(  \mathbf{k}\right)
-\frac{1}{2}\mathfrak{z}^{\ast}\left(  \mathbf{k}\right)  \partial_{t}%
\mathfrak{Z}\left(  \mathbf{k}\right)  \right.  \nonumber\\
& +\frac{1}{2}\mathfrak{Z}\left(  \mathbf{k}\right)  \partial_{t}\mathfrak{z}^{\ast
}\left(  \mathbf{k}\right)  +\frac{1}{2}\mathfrak{z}\left(  \mathbf{k}\right)
\partial_{t}\mathfrak{Z}^{\ast}\left(  \mathbf{k}\right)  -i\Omega\left(
\mathfrak{Z}^{\ast}\left(  \mathbf{k}\right)  \mathfrak{z}\left(  \mathbf{k}\right)
+\mathfrak{z}^{\ast}\left(  \mathbf{k}\right)  \mathfrak{Z}\left(  \mathbf{k}\right)
\right)  \nonumber\\
& \left.  -iK\left(  \mathfrak{z}^{\ast}\left(  \mathbf{k}\right)  \mathfrak{Z}^{\ast
}\left(  -\mathbf{k}\right)  +\mathfrak{z}\left(  \mathbf{k}\right)
\mathfrak{Z}\left(  -\mathbf{k}\right)  \right)  \right]  \nonumber\\
& +J^{\prime^{2}}\sum_{\mathbf{k},\mathbf{k}^{\prime}}\int dtdt^{\prime
}e^{-i\varepsilon_{1,2}\left(  \mathbf{k},\mathbf{k}^{\prime}\right)  \left(
t-t^{\prime}\right)  }\left[  \mathfrak{Z}^{\ast}\left(  t,\mathbf{p}\right)
\mathcal{U}_{12}\left(  t-t^{\prime}\right)  \mathfrak{z}\left(  t^{\prime
},-\mathbf{p}\right)  \right.  \nonumber\\
& \left.  +\mathfrak{z}^{\ast}\left(  t,\mathbf{p}\right)  \mathcal{U}_{21}\left(
t-t^{\prime}\right)  \mathfrak{Z}\left(  t,-\mathbf{p}\right)  +\mathfrak{z}^{\ast
}\left(  t,\mathbf{p}\right)  \mathcal{U}_{22}\left(  t,t^{\prime}\right)
\mathfrak{z}\left(  t^{\prime},-\mathbf{p}\right)  \right]  \nonumber\\
& +i\sum_{\mathbf{k}}\int dt\left(  Q_{c}^{\ast}\mathfrak{Z}+Q_{c}\mathfrak{Z}^{\ast
}\right)  +i\sum_{\mathbf{k}}\int dt\left(  J_{c}^{\ast}\mathfrak{z}+J_{c}%
\mathfrak{z}^{\ast}\right)  ,
\end{align}
where we have $\mathbf{p}  =\mathbf{k}^{\prime}-\mathbf{k}$ and $\varepsilon_{12}\left(  \mathbf{k,k}^{\prime}\right)  =\varepsilon
_{1}\left(  \mathbf{k}\right)  -\varepsilon_{2}\left(  \mathbf{k}^{\prime
}\right)  $. The remaining variables are defined as follows:
\begin{align}
\mathfrak{Z}  & =\frac{1}{\sqrt{2}}\left(  S_{x}+iS_{y}\right)  ,\\
\mathfrak{z}  & =\frac{1}{\sqrt{2}}\left(  D_{x}+iD_{y}\right), \\
Q_{c}  & =\frac{1}{\sqrt{2}}\left(  Q_{x}+iQ_{y}\right) ,  \\
J_{c}  & =\frac{1}{\sqrt{2}}\left(  J_{x}+iJ_{y}\right) .
\end{align}
Varying the action with respect to its variables, we get the respective
equations of motion for the magnetization and the 
fluctuations. \ A solution of these 
equations
will require a careful treatment of the boundary conditions. \ To do this, 
we found it easier to take a related representation of the coherent states. \ This 
representation is equivalent to the usual harmonic oscillator representation.

 We  define two harmonic oscillator-type operators $a$ and $a^{+}$
such that

\begin{align}
a  &  =\frac{\sqrt{2}}{2}\left[  \left(  \frac{c_{xx}}{c_{yy}}\right)
^{1/4}\widehat{S}_{x}+i\left(  \frac{c_{yy}}{c_{xx}}\right)  ^{1/4}\widehat
{S}_{y}\right]  ,\nonumber\\
a^{+}  &  =\frac{\sqrt{2}}{2}\left[  \left(  \frac{c_{xx}}{c_{yy}}\right)
^{1/4}\widehat{S}_{x}-i\left(  \frac{c_{yy}}{c_{xx}}\right)  ^{1/4}\widehat
{S}_{y}\right]  .
\end{align}
Hence, the energy operator for the spin system becomes
\begin{equation}
\widehat{H}=\hbar\omega_{0}\left(  a^{+}a+\frac{1}{2}\right)  .
\end{equation}
with the frequency given by
\begin{equation}
\omega_{0}=S\sqrt{c_{xx}c_{yy}}\; ; \;\;\;\; (S=1).
\end{equation}
Therefore coherent states in this representation are defined by
\begin{equation}
a\left|  Z\right\rangle =Z\left|  Z\right\rangle .
\end{equation}
To change between representations $\left(  \mathfrak{Z,z}\right)  \rightarrow
\left(  Z,z\right)  $ ( with careful handling of the boundary conditions) in
the path integral representations we need to make the change of variables%
\begin{align}
\mathfrak{Z}  &  \rightarrow\frac{1}{2}\left(  c_{1}Z+c_{2}Z^{\ast}\right)
,\nonumber\\
\mathfrak{z}  &  \rightarrow\frac{1}{2}\left(  c_{1}z+c_{2}z^{\ast}\right)  ,
\end{align}
with the coefficients $c_{1}$ and $c_{2}$ given by
\begin{align}
c_{1}  &  =\frac{\sqrt{c_{xx}}+\sqrt{c_{yy}}}{2\sqrt{\omega_{0}}},\nonumber\\
c_{2}  &  =-\frac{\sqrt{c_{xx}}-\sqrt{c_{yy}}}{2\sqrt{\omega_{0}}}.
\end{align}
It will be seen that in the high frequency regime, the damping 
depends separately on $c_1$ and $c_2$. 
\ The path integral representation on the closed-time path requires doubling of
variables as has been done above. The following boundary conditions are also needed:
\begin{align}
Z_{1}\left(  t_{f}\right)   &  =Z_{2}\left(  t_{f}\right)  ,\label{bc1}\\
Z_{1}\left(  t_{0}\right)   &  =\exp\left[  -\beta\omega_{0}\right]
Z_{2}\left(  t_{0}\right) , \\
Z_{1}^{\ast}\left(  t_{0}\right)   &  =\exp\left[ +\beta\omega_{0}\right]
Z_{2}^{\ast}\left(  t_{0}\right).
\end{align}
It is important to observe that the boundary conditions are 
non-hermitian in the coherent-state formulation. \ This is a 
direct result of the application of the Kubo-Martin-Schwinger (KMS)
condition. \ This 
 implies that $Z$, 
the average sum of $Z_{1}$ and $Z_{2}$ ,  \ and their respective
fluctuations $z$ satisfy the conditions
\begin{equation}
Z\left(  t_{0}\right)  =-\frac{1}{2}\coth\left(  \beta\frac{\omega_{0}}%
{2}\right)  z\left(  t_{0}\right)  ,
\end{equation}
and
\begin{equation}
z\left(  t_{f}\right)  =0. \label{bc2}
\end{equation}
The free action in the new representation therefore takes the form
\begin{align}
iS^{\left(  0\right)  }\left[  Z^{\ast},Z,z^{\ast},z\right]   &  =\frac{1}%
{2}\left(  Z_{f}Z_{f}^{\ast}\left(  t_{f}\right)  +Z_{f}^{\ast}Z_{f}\left(
t_{f}\right)  \right)  +\frac{1}{2}\left(  Z_{i}Z^{\ast}\left(  t_{0}\right)
+Z_{i}^{\ast}Z\left(  t_{0}\right)  \right) \nonumber\\
&  +\frac{1}{4}\left(  Z_{i}z^{\ast}\left(  t_{0}\right)  +Z_{i}^{\ast
}z\left(  t_{0}\right)  \right) \nonumber\\
&  +i\int_{t_{0}}^{t_{f}}dt\left\{  \frac{1}{2i}\left(  \overset{\cdot
}{Z^{\ast}}z-Z^{\ast}\overset{\cdot}{z}\right)  +\frac{1}{2i}\left(
\overset{\cdot}{z^{\ast}}Z-z^{\ast}\overset{\cdot}{Z}\right)  \right.
\nonumber\\
&  \left.  -\omega_{0}\left(  z^{\ast}Z+zZ^{\ast}\right)  +Z^{\ast}%
\Delta+Z\Delta^{\ast}+z^{\ast}\Sigma+z\Sigma^{\ast}\right\}  ,
\end{align}
where the external sources $\Delta$, $\Sigma$ are defined through the relation%

\begin{equation}
\mathbf{J}_{1}\cdot\mathbf{S}_{1}-\mathbf{J}_{2}\cdot\mathbf{S}_{2}=Z^{\ast
}\Delta+Z\Delta^{\ast}+z^{\ast}\Sigma+z\Sigma^{\ast}.
\end{equation}
\ The equations of motion are obtained from the full action
$iS^{eff}\left[  Z^{\ast},Z,z^{\ast},z\right]  =iS^{\left(  0\right)  }\left[
Z^{\ast},Z,z^{\ast},z\right]  +iS^{B}\left[  Z^{\ast},Z,z^{\ast},z\right]  $,
where the last term is due to the interaction of the conduction electrons with
the magnetization. \  For the fluctuations $z$ and $z^{\ast}$, we obtain

\begin{eqnarray}
\left(  \partial_{t}+i\omega_{0}\right)  z\left(  t,\mathbf{p}\right)&
-\frac{J^{2}}{4}\int dt^{\prime}G_{1}\left(  t-t^{\prime},\mathbf{p}%
\right)  z\left(  t^{\prime},-\mathbf{p}\right) \nonumber \\
&-\frac{J^{2}}{4}\int
dt^{\prime}G_{2}\left(  t-t^{\prime},\mathbf{p}\right)  z^{\ast}\left(
t^{\prime},-\mathbf{p}\right)  \nonumber \\
 = i \Delta\left(  t,\mathbf{p}\right)  ,
\end{eqnarray}
and
\begin{eqnarray}
\left(  \partial_{t}-i\omega_{0}\right)  z^{\ast}\left(  t,\mathbf{p}\right) &
+\frac{J^{2}}{4}\int dt^{\prime}G_{3}\left(  t-t^{\prime},\mathbf{p}%
\right)  z^{\ast}\left(  t^{\prime},-\mathbf{p}\right) \nonumber \\ 
& +\frac{J^{2}}%
{4}\int dt^{\prime}G_{2}\left(  t-t^{\prime},\mathbf{p}\right)  z^{\ast
}\left(  t^{\prime},-\mathbf{p}\right) \nonumber \\
  = -i \Delta^{\ast}\left(  t,\mathbf{p}
\right)  . 
\end{eqnarray}
To solve these equations and take account of 
the finite size of the film, we introduce the following definition for the 
'averaged' Green functions,
\begin{equation}
\overline{G}_{\alpha}\left(  t-t^{\prime}\right)  =\frac{1}{D A}\int d^2 x_\parallel 
\int d^2 y_\parallel \int
_{-D/2}^{D/2}dxe^{ipx}\int_{-D/2}^{D/2}dye^{ipy}\int\frac{d^{3}p}{\left(
2\pi\right)  ^{3}}G_{\alpha}\left(  t-t^{\prime},\mathbf{p}\right).
\end{equation}
where $A$ is the area of the surface of the film. \ This definition is useful 
for solving for the zero mode only. \ As we did in the previous section, this 
is obtained by  Fourier transforming back the equations of motion 
to the real space representation where the condition of finite thickness 
can be easily implemented. \ These Green functions $(\alpha = 1, .., 5)$ are 
given explicitly in the 
appendix. \ It should be noted that the Green functions $\overline{G}_\alpha$ with $\alpha = 1,2,3$  are directly 
related to the appearance of the dissipative term in 
the magnetization. \ They involve terms 
similar to those that appeared in Eq. \ref{f2f1}, but they 
are also the Green functions that appear in the 
equations of motion of the fluctuations $z$ and $z^\ast$. \ Hence 
their dissipative nature is very clear in this formalism since 
they introduce irreversibility in the dynamics
 of $Z$ and $Z^\ast$.

\ The corresponding Green's function needed for the solution of the
fluctuations are therefore given by
\begin{align}
\left(  \partial_{t}+i\omega_{0}\right)  \mathfrak{g}_{1}\left(  t-t^{\prime
}\right)  -\frac{J^{2}}{4}\int dt^{\prime\prime}\overline{G}_{1}\left(
t-t^{\prime\prime}\right)  \mathfrak{g}_{1}\left(  t^{\prime\prime}-t^{\prime
}\right)    & =\delta\left(  t-t^{\prime}\right)  ,\label{g1}\\
\mathfrak{g}_{1}\left(  t -t^{\prime}\right)    & =0; \; \; \; \; t>t^{\prime
}
\end{align}
and
\begin{align}
\left(  \partial_{t}+i\omega_{0}\right)  \mathfrak{g}_{1}^{^{\prime}}\left(
t-t^{\prime}\right)  +\frac{J^{{2}}}{4}\int d\tau\overline{G}%
_{1}\left(  t-\tau\right)  \mathfrak{g}_{1}^{^{\prime}}\left(  \tau-t^{\prime
}\right)    & =\delta\left(  t-t^{\prime}\right)  ,\\
\mathfrak{g}_{1}^{^{\prime}}\left(  t-t^{\prime}\right)    & =0;\; \; \; \; \; t<t^{\prime}%
\end{align}
\ Similarly, we get two  more equations for $Z(t)$ and $Z^{\ast}(t)$ that 
involve two more Green functions. \ We only write the respective solutions 
below.\ Therefore, the solutions of the fluctuations and the average sum
take the form

\begin{align}
z\left(  t\right)   &  =i\int\mathfrak{g}_{1}\left(  t-t^{\prime}\right)
\Delta\left(  t^{^{\prime}}\right)  dt^{^{\prime}}\nonumber\\
&  -i\left(  \frac{J}{2}\right)  ^{2}\int d\tau dt^{^{^{\prime}}%
}dt^{^{\prime\prime}}\mathfrak{g}_{1}\left(  t-t^{^{\prime}}\right)  \overline
{G}_{2}\left(  t-t^{^{\prime\prime}}\right)  \mathfrak{g}_{1}^{\ast}\left(
t^{^{\prime\prime}}-\tau\right)  \Delta^{\ast}\left(  \tau\right),
\end{align}
and
\begin{align}
Z\left(  t\right)    & =i\int dt^{\prime}\mathfrak{g}_{1}^{\prime}\left(
t-t^{^{\prime}}\right)  \Sigma\left(  t^{\prime}\right)  +i\frac{J^2}
{4}\int d\tau d\tau^{\prime}dt^{\prime}\mathfrak{g}_{1}^{\prime}\left(
t-\tau\right)  \overline{G}_{4}\left(  \tau-t^{\prime}\right)  \mathfrak{g}%
_{1}\left(  t^{\prime}-\tau\right)  \Delta\left(  \tau^{\prime}\right)
\nonumber\\
& -i\frac{J^{2}}{4}\int d\tau d\tau^{\prime}dt^{\prime}\mathfrak{g}
_{1}^{\prime}\left(  t-\tau\right)  \overline{G}_{5}\left(  \tau-t^{\prime
}\right)  \mathfrak{g}_{1}^{\ast}\left(  t^{\prime}-\tau\right)  \Delta^{\ast
}\left(  \tau^{\prime}\right)  \nonumber\\
& -i\frac{J^{2}}{4}\int d\tau d\tau^{\prime}dt^{\prime}\mathfrak{g}%
_{1}^{\prime}\left(  t-\tau\right)  \overline{G}_{2}\left(  \tau-t^{\prime
}\right)  \mathfrak{g}_{1}^{\prime\ast}\left(  t^{\prime}-\tau\right)
\Sigma^{\ast}\left(  \tau^{\prime}\right).
\end{align}
Since $z\left(  t_{f}\right)  =0$, we require that $\mathfrak{g}_{1}\left(
t-t^{\prime}\right)  =0$\ for $t>t^{\prime}$. 

First, we recalculate the initial correlations to show that we have the
correct boundary conditions. The initial value for $Z$ follows from the
solution for $Z(t)$ after setting the coupling constant $J=0$
\begin{equation}
Z\left(  t_{0}\right)  =-\frac{1}{2}\coth\left(  \beta\frac{\omega_{0}}%
{2}\right)  i\int dt^{\prime}\overline{g}_{1}\left(  t_{0}-t^{\prime}\right)
\Delta\left(  t^{\prime}\right)  .
\end{equation}
Then the derivative of the x-component of the magnetization 
with respect of the external sources is 
\begin{equation}
\frac{1}{i}\frac{\delta\widehat{S}_{x}\left(  t\right)  }{\delta\Delta
J_{x}\left(  t^{\prime}\right)  }=-\frac{1}{2^{2}}\frac{c_{yy}}{\omega_{0}%
}\coth\left(  \beta\frac{\omega_{0}}{2}\right)  \left[  \mathfrak{g}_{1}\left(
t-t^{\prime}\right)  +\mathfrak{g}_{1}^{\ast}\left(  t-t^{\prime}\right)
\right]
\end{equation}
In the free theory, the propagator $\mathfrak{g}_{1}$ is simply given by%
\begin{equation}
\mathfrak{g}_{1}\left(  t-t^{\prime}\right)  =-\Theta\left(  t^{\prime}-t\right)
e^{-i\omega_{0}\left(  t-t^{\prime}\right)  }.
\end{equation}
Hence
\begin{align}
\frac{1}{i}\frac{\delta\left\langle \widehat{S}_{x}\right\rangle }%
{\delta\Delta J_{x}\left(  t\right)  } &  =\frac{1}{2}\frac{c_{yy}}{\omega
_{0}}\coth\left(  \beta\frac{\omega_{0}}{2}\right)  \cos\omega_{0}t\nonumber\\
&  =\frac{1}{2}\left\langle \left\{  S_{x},S_{x}\left(  t\right)  \right\}
\right\rangle ,
\end{align}
which is the desired relation that was derived in Sect. II by 
a different method. \ To get this solution, it was crucial that we 
apply the correct boundary conditions on $Z$ and $z$, Eqs. \ref{bc1}-\ref{bc2}.

\ For the coupled case, the initial condition for the $Z(t)$ equation is
\begin{align}
Z\left(  t_{0}\right)   &  =-\frac{1}{2}\coth\left(  \beta\frac{\omega_{0}}%
{2}\right)  \left\{  -i\int_{t_{0}}^{\infty}dt^{\prime}\mathfrak{g}_{1}\left(
t_{0}-t^{\prime}\right)  \Delta\left(  t^{\prime}\right)  \right.  \nonumber\\
&  -i\left(  \frac{J}{2}\right)  ^{2}\int_{t_{0}}^{\infty}d\tau
dt^{^{\prime}}dt^{\prime\prime}\mathfrak{g}_{1}\left(  t-t^{\prime}\right)
\overline{G}_{2}\left(  t-t^{"}\right)  \mathfrak{g}_{1}^{\ast}\left(
t^{\prime\prime}-\tau\right)  \Delta^{\ast}\left(  \tau\right)  .
\end{align}
Hence the general solution for $Z(t)$ is
\begin{align}
Z\left(  t\right)   &  =Z_{0}\left(  t\right)  +i\int_{t_{0}}^{\infty}%
d\tau\mathfrak{g}_{1}^{^{\prime}}\left(  t-\tau\right)  \Sigma\left(  \tau\right)
\\
&  +i\left(  \frac{J}{2}\right)  ^{2}\int_{t_{0}}^{\infty}d\tau
d\tau^{^{\prime}}dt^{\prime}\mathfrak{g}_{1}^{^{\prime}}\left(  t-\tau^{\prime
}\right)  \overline{G}_{4}\left(  t-\tau\right)  \mathfrak{g}_{1}\left(
t^{\prime}-\tau^{\prime}\right)  \Delta\left(  \tau^{\prime}\right)
\nonumber\\
&  -i\left(  \frac{J}{2}\right)  ^{2}\int_{t_{0}}^{\infty}d\tau
d\tau^{^{\prime}}dt^{\prime}\mathfrak{g}_{1}^{^{\prime}}\left(  t-\tau\right)
\overline{G}_{5}\left(  \tau-t^{\prime}\right)  \mathfrak{g}_{1}^{\ast}\left(
t^{\prime}-\tau^{\prime}\right)  \Delta^{\ast}\left(  \tau^{\prime}\right)
\nonumber\\
&  -i\left(  \frac{J}{2}\right)  ^{2}\int_{t_{0}}^{\infty}d\tau
d\tau^{^{\prime}}dt^{\prime}\mathfrak{g}_{1}^{^{\prime}}\left(  t-\tau\right)
\overline{G}_{2}\left(  \tau-t^{\prime}\right)  \mathfrak{g}_{1}^{\ast ^\prime}\left(
t^{\prime}-\tau^{\prime}\right)  {\Sigma}^{\star}\left(  \tau^{\prime
}\right)
\end{align}
where \ $Z_{0}\left(  t\right)  $ is a particular solution of the
nonhomogeneous problem

\begin{align}
Z_{0}\left(  t\right)    & =Z_{0}\left(  t\right)  e^{-i\omega_{0}\left(
t-t_{0}\right)  }+\left(  \frac{J}{2}\right)  ^{2}e^{-i\omega_{0}%
t}\int_{t_{0}}^{t}dt^{\prime}e^{i\omega_{0}t^{\prime}}\left[  -\int
dt^{^{\prime\prime}}\overline{G}_{1}\left(  t^{^{\prime}}-t^{^{\prime\prime}%
}\right)  e^{-i\omega_{0}\left(  t^{^{\prime\prime}}-t_{0}\right)  }%
Z_{0}\left(  t_{0}\right)  \right.  \nonumber\\
& \left.  \int dt^{^{\prime\prime}}\overline{G}_{2}\left(  t^{\prime
}-t^{^{\prime\prime}}\right)  e^{i\omega_{0}\left(  t^{^{\prime\prime}}%
-t_{0}\right)  }\overline{Z}_{0}\left(  t_{0}\right)  \right]
\end{align}
\ Similarly, we get the solution for $Z^{\ast}(t)$
\begin{align}
Z^{\ast}\left(  t\right)   &  =Z_{0}^{\ast}\left(  t_{0}\right)
e^{i\omega_{0}\left(  t-t_{0}\right)  }-i\int_{t_{0}}^{\infty}dt^{\prime
}\mathfrak{g}_{1}^{\ast^{\prime}}\left(  t-t^{\prime}\right)  \Sigma^{\ast}\left(
t^{\prime}\right)  \nonumber\\
&  -i\left(  \frac{J}{2}\right)  ^{2}\int d\tau d\tau^{^{\prime}
}dt^{\prime}\mathfrak{g}_{1}^{^{\prime}\ast}\left(  t-\tau\right)  \overline
{G}_{2}\left(  \tau-t^{\prime}\right)  \mathfrak{g}_{1}^{^{\prime}}\left(
t^{\prime}-\tau^{\prime}\right)  \Sigma\left(  \tau^{\prime}\right)
\nonumber\\
&  -i\left(  \frac{J}{2}\right)  ^{2}\int_{t_{0}}d\tau d\tau
^{^{\prime}}dt^{\prime}\mathfrak{g}_{1}^{^{\prime}\ast}\left(  t-\tau\right)
\overline{G}_{5}\left(  \tau-t^{\prime}\right)  \mathfrak{g}_{1}\left(  t^{\prime
}-\tau^{\prime}\right)  \Delta\left(  \tau^{\prime}\right)  \nonumber\\
&  +i\left(  \frac{J}{2}\right)  ^{2}\int_{t_{0}}^{\infty}d\tau
d\tau^{^{\prime}}dt^{\prime}\mathfrak{g}_{1}^{^{\prime}}\left(  t-\tau\right)
\overline{G}_{4}\left(  \tau-t^{\prime}\right)  \mathfrak{g}_{1}^{\ast
}\left(  t^{\prime}-\tau^{\prime}\right)  \Delta^{\ast}\left(  \tau^{\prime
}\right)  \nonumber\\
&  +\left(  \frac{J}{2}\right)  ^{2}e^{i\omega_{0}\left(
t-t_{0}\right)  }\int_{t_{0}}^{t}dt^{\prime}e^{-i\omega_{0}\left(
t^{^{\prime}}-t_{0}\right)  }\left[  \int dt^{^{\prime\prime}}\overline{G}%
_{3}\left(  t-t^{^{\prime\prime}}\right)  e^{i\omega_{0}\left(  t^{^{\prime
\prime}}-t_{0}\right)  }Z_{0}^{\ast}\left(  t_{0}\right)  \right.  \nonumber\\
&  \left.  -\int dt^{^{\prime\prime}}\overline{G}_{2}\left(  t^{\prime
}-t^{^{\prime\prime}}\right)  e^{-i\omega_{0}\left(  t^{^{\prime\prime}}
-t_{0}\right)  }Z_{0}\left(  t_{0}\right)  \right]  ,
\end{align}
with the initial state given by%
\begin{equation}
Z_{0}\left(  t_{0}\right)  =-\frac{1}{2}i\coth\left(  \beta\frac{\omega_{0}%
}{2}\right)  \int dt^{\prime}\mathfrak{g}_{1}\left(  t_{0}-t^{\prime}\right)
\Delta\left(  t^{\prime}\right)  .
\end{equation}
\ Now, it is easy to calculate the components of the magnetization from
the above results. We just need to differentiate the average magnetization
with respect to $\left(  j_{x},j_{y}\right)  $,

\begin{align}
j_{x}  & =\frac{\sqrt{2}}{2}\sqrt{\frac{\omega_{0}}{c_{yy}}}\left(
\Delta+\Delta^{\ast}\right)  ,\\
j_{y}  & =\frac{\sqrt{2}}{2i}\sqrt{\frac{\omega_{0}}{c_{xx}}}\left(
\Delta-\Delta^{\ast}\right),
\end{align}
 to find the symmetric
correlation functions for the magnetization.

\ A  straightforward calculation gives the differential of the
x-component with respect to $j_{x}$

\begin{align}
\frac{1}{i}\frac{\delta\left\langle \widehat{S}_{x}(t)\right\rangle }{\delta
j_{x}\left(  t^{\prime}\right)  } &  =\frac{1}{2}\frac{c_{yy}}
{\omega_{0}}\left\{  -\frac{1}{2}\coth\left(  \beta\frac{\omega_{0}}{2}\right)  \left[
e^{i\omega_{0}\left(  t-t_{0}\right)  }\mathfrak{g}_{1}^{\ast}\left(
t_{0}-t^{\prime}\right) + e^{-i\omega_{0}\left(  t-t_{0}\right)  }\mathfrak{g}%
_{1}\left(  t_{0}-t^{\prime}\right)  \right]  \right. \nonumber \\
&  +\left(  \frac{J}{2}\right)  ^{2}\frac{1}{2}\coth\left(
\beta\frac{\omega_{0}}{2}\right)  \left[  e^{-i\omega_{0}\left(
t-t_{0}\right)  }\int_{t_{0}}^{t}d\tau\int d\tau^{\prime}e^{i\omega_{0}\left(
\tau-t_{0}\right)  }\right.  \nonumber  \\
&  \times\left[  \overline{G}_{1}\left(  \tau-\tau^{\prime}\right)
e^{-i\omega_{0}\left(  \tau^{\prime}-t_{0}\right)  }\mathfrak{g}_{1}\left(
t_{0}-t^{\prime}\right)  - \overline{G}_{2}\left(  \tau-\tau^{\prime}\right)
e^{i\omega_{0}\left(  \tau^{\prime}-t_{0}\right)  }\mathfrak{g}_{1}^{\ast}\left(
t_{0}-t^{\prime}\right)  \right]\nonumber  \\
&  - e^{i\omega_{0}\left(  t-t_{0}\right)  }\int_{t_{0}}^{t}d\tau\int
d\tau^{\prime}e^{-i\omega_{0}\left(  \tau-t_{0}\right)  }\left[  \overline
{G}_{3}\left(  \tau-\tau^{\prime}\right)  e^{i\omega_{0}\left(  \tau^{\prime
}-t_{0}\right)  }\mathfrak{g}_{1}^{\ast}\left(  t_{0}-t^{\prime}\right)  \right.\nonumber \\
&  \left.  \left.   -\overline{G}_{2}\left(\tau-\tau^{\prime}\right)
e^{-i\omega_{0}\left(  \tau^{\prime}-t_{0}\right)  }\mathfrak{g}_{1}\left(
t_{0}-t^{\prime}\right)  \right]  \right]   \nonumber \\
&  +\left(  \frac{J}{2}\right)  ^{2}\left[  \int d\tau d\tau
^{^{\prime}}\left[  \mathfrak{g}_{1}^{\prime}\left(  t-\tau\right)  \overline{G}_{4}\left(
\tau-\tau^{\prime}\right)  \mathfrak{g}_{1}\left(  \tau^{\prime}-t^{\prime
}\right)  \right.  \right.  \nonumber  \\
&  -\mathfrak{g}_{1}^{\ast^\prime}\left(  t-\tau\right)  \overline{G}_{5}\left(
\tau-\tau^{\prime}\right)  \mathfrak{g}_{1}\left( \tau^{\prime}-t^{\prime
}\right)  -\mathfrak{g}_{1}^{^{\prime}}\left(  t-\tau\right)  \overline{G}%
_{5}\left(  \tau-\tau^{\prime}\right)  \mathfrak{g}_{1}^{\ast}\left(  \tau
^{\prime}-t^{\prime}\right)  \nonumber \\
&  +\left.  \left.  \left. \mathfrak{g}_{1}^{\ast^\prime}\left(  t-\tau\right)
\overline{G}_{4}\left(  \tau-\tau^{\prime}\right)  \mathfrak{g}
_{1}^{\ast}\left(  \tau^{\prime}-t^{\prime}\right)  \right]  
\right] \right\} \nonumber  \\
& = \frac{1}{2}\langle \left\{\widehat{S}_{x}(t),  \widehat{S}_{x}(t') 
\right\} \rangle.
\end{align}
\ This is a general result for the noise due to 
spin-flip scattering between magnons and 
conduction electrons that is useful 
for all frequencies. \ It is clear from this expression, that 
the correlation functions 
depend explicitly on the initial state of the system. \ However, 
there is a term which is  invariant under time translation. \ This 
part of the correlation term survives at times much later than 
the initial conditions. \ Before we turn to the calculation of 
this term which is the term usually measured in FMR-type experiments 
we give the expression for the
 correlation function of $S_x$ and $S_y$ components of the 
magnetization
\begin{align}
\frac{1}{i}\frac{\delta\left\langle \widehat{S}_{x}\left(  t\right)
\right\rangle }{\delta j_{y}\left(  t^{\prime}\right)  } &  =\frac{i}%
{2}\left\{  -\frac{1}{2}\coth\left(  \beta\frac{\omega_{0}}{2}\right)  \left[
e^{-i\omega_{0}\left(  t-t_{0}\right)  }\mathfrak{g}_{1}\left(  t_{0}-t^{\prime
}\right)  - e^{i\omega_{0}\left(  t-t_{0}\right)  }\mathfrak{g}_{1}^{\ast}\left(
t_{0}-t^{\prime}\right)  \right]  \right.  \nonumber\\
&  +\left(  \frac{J}{2}\right)  ^{2}\left[  \int d\tau d\tau
^{^{\prime}}\left[  \mathfrak{g}_{1}^{\prime}\left(  t-\tau\right)
\overline{G}_{4}\left(  \tau-\tau^{\prime}\right)  \mathfrak{g}_{1}\left(
\tau^{\prime}-t^{\prime}\right)  \right.  \right.  \nonumber\\
&  -\mathfrak{g}_{1}^{\ast^{\prime}}\left(  t-\tau\right)  \overline{G}_{5}\left(
\tau-\tau^{\prime}\right)  \mathfrak{g}_{1}\left(  \tau^{\prime}-t^{\prime
}\right)  +\mathfrak{g}_{1}^{^{\prime}}\left(  t-\tau\right)  \overline{G}%
_{5}\left(  \tau-\tau^{\prime}\right)  \mathfrak{g}_{1}^{\ast}\left(  \tau
^{\prime}-t^{\prime}\right)  \nonumber\\
&  \left.  - \mathfrak{g}_{1}^{\ast^\prime}\left(  t-\tau\right)  \overline{G}%
_{4}\left(  \tau-\tau^{\prime}\right)  \mathfrak{g}_{1}^{\ast}\left(
\tau^{\prime}-t^{\prime}\right)  \right]  \nonumber\\
&  +\frac{i}{2}\left(  \frac{J}{2}\right)  ^{2}e^{-i\omega_{0}\left(
t-t_{0}\right)  }\coth\left(  \beta\frac{\omega_{0}}{2}\right)  \int_{t_{0}%
}^{t}d\tau e^{i\omega_{0}\left(  \tau-t_{0}\right)  }\nonumber\\
&  \times   \int dt^{^{\prime\prime}}\left[  \overline{G}_{1}\left(
\tau-t^{^{\prime\prime}}\right)  e^{-i\omega_{0}\left(  t^{\prime\prime}%
-t_{0}\right)  }\mathfrak{g}_{1}\left(  t_{0}-t^{\prime}\right)
+\overline{G}_{2}\left(  \tau-t^{^{\prime\prime}}\right)  e^{i\omega
_{0}\left(  t^{\prime\prime}-t_{0}\right)  }\mathfrak{g}_{1}^{\ast}\left(
t_{0}-t^{\prime}\right)  \right]   \nonumber\\
&  +\frac{i}{2}\left(  \frac{J}{2}\right)  ^{2}e^{i\omega_{0}\left(
t-t_{0}\right)  }\coth\left(  \beta\frac{\omega_{0}}{2}\right)  \int_{t_{0}%
}^{t}d\tau e^{-i\omega_{0}\left(  \tau-t_{0}\right)  }\nonumber\\
&  \times  \left.  \int dt^{^{\prime\prime}} \left[ \overline{G}_{3}\left(
\tau-t^{^{\prime\prime}}\right)  e^{i\omega_{0}\left(  t^{\prime\prime}%
-t_{0}\right)  }\mathfrak{g}_{1}^{\ast}\left(  t_{0}-t^{\prime}\right)  +\int
dt^{^{\prime\prime}}\overline{G}_{2}\left(  \tau-t^{^{\prime\prime}}\right)
e^{-i\omega_{0}\left(  t^{\prime\prime}-t_{0}\right)  }\mathfrak{g}_{1}\left(
t_{0}-t^{\prime}\right)  \right]  \right\}  .
\end{align}
\ An explicit expression for these correlation functions is not 
needed for what follows,  but we will write its limit in the 
adiabatic limit which is the case of most current interest for 
the average magnetization of the film. \ Before we 
do that, we would like to make few more comments about these general 
expressions for the correlation functions. \ These expressions are beyond 
the usual fluctuation-dissipation relation and hence they can be adapted
to truly 'non-equilibrium' situations. \ As an 
example we mention  time-dependent
pulse-field excitations of the magnetization, switching by a magnetic
field. \ In this latter case, the z-axis is the local equilibrium 
axis. \ They also give us an idea on the noise behavior in 
local magnetic moments and the corresponding damping.

\bigskip

\begin{figure}[ht]
  \begin{center}
 \mbox{\epsfig{file=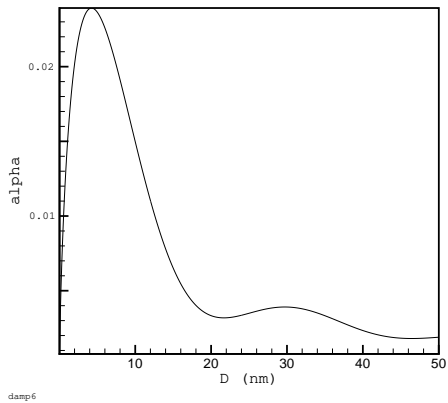,height=6 cm}}
  \end{center}
  \caption{{{The damping as a function of the thickness for
 $\delta_r = 0.2$  and $ c_{xx} = 100 c_{yy}$. The last term is 
a measure of anisotropy and is on the
high  side for realistic permalloy films.}}}
  \label{damp6}
\end{figure}

\begin{figure}[ht]
  \begin{center}
 \mbox{\epsfig{file=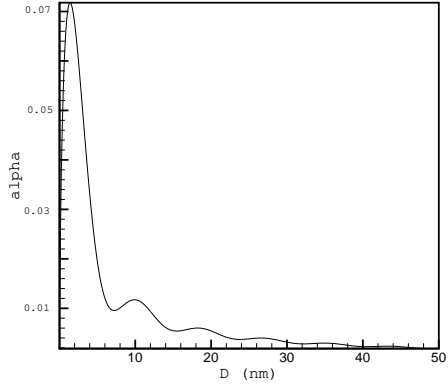,height=6 cm}}
  \end{center}
  \caption{{{The damping as a function of the thickness
 for $\delta_r = 0.6$ and $ c_{xx} = 100 c_{yy}$ (same as fig. 
\ref{damp6}).  }}}
  \label{damp7}
\end{figure}

To calculate the noise spectrum in the x-component of the magnetization we
need the Fourier component of the function
\begin{equation}
C_{xx}\left(  t-t^{\prime}\right)  =\left.  \left\langle S_{x}\left(
t\right)  S_{x}\left(  t^{\prime}\right)  \right\rangle \right|
_{t>>t_{0},t^{\prime}>>t_{0}},
\end{equation}
\ since here we will not address the transient regime, which is 
treated elsewhere \cite{rebei_ox}. \ Its 
Fourier transform is easily found to be

\begin{eqnarray}
C_{xx}\left(  \omega\right)    =\frac{c_{yy}}{\omega_{0}} \left(
\frac{J}{2}\right)^{2} \Re   \left\{   \mathfrak{g}_{1}^{\prime}
\left(  \omega\right)
\overline{G}_{4}\left(  \omega\right)  \mathfrak{g}_{1}\left(  \omega\right) 
-\mathfrak{g}_{1}^{\prime}\left(  \omega\right)
\overline{G}_{5}\left(  \omega\right)  \mathfrak{g}_{1}^{\ast}\left(-\omega\right) \right\}.
\label{noisexx}
\end{eqnarray}
\ The damping for large thicknesses $D$ (and nonzero exchange splitting)
acquires a simple asymptotic expression,
\begin{equation}
\alpha \approx \frac{\pi\left(  c_{1}^{2}+c_{2}^{2}\right)  }{2^{5}\left(
k_{F}^{\uparrow}+k_{F}^{\downarrow}\right)  D}\left(  1+2\frac{\sin\left(
\frac{1}{4}\left(  k_{F}^{\uparrow}-k_{F}^{\downarrow}\right)  D\right)
}{\left(  k_{F}^{\uparrow}-k_{F}^{\downarrow}\right)  D}+48\frac{\cos\left(
\frac{1}{4}\left(  k_{F}^{\uparrow}-k_{F}^{\downarrow}\right)  D\right)
}{\left(  k_{F}^{\uparrow}-k_{F}^{\downarrow}\right)  ^{2}D^{2}} + ...\right)
\end{equation}
This damping differs from the one found in the previous sections in two 
different aspects. \ The first is that for large thicknesses, the damping 
is weakly dependent on the sd-exchange energy $\delta_r$ as opposed 
to being linear in $\delta_r$. \ This result is now much similar to 
Mills and Berger. \ The second important difference is that the 
relaxation time depends explicitly on the symmetry of the original Hamiltonian 
of the magnetization. \ This result is however
similar  to what we found in ref. 
\onlinecite{rebei}. \ For circular precession, we simply have 
$c_1^2+c_2^2 = 1$ and hence any dependence on the form of 
the precession is lost. \ The damping is still however 
a scalar of the Gilbert form for $\omega << \delta/\hbar$ and does 
not appear to require 
a tensor form as suggested in ref. 
\onlinecite{safonov}. \ The high frequency regime, which 
is applicable to atomic moments, is however more interesting in 
this respect. \ The relaxation time is not a simple function 
of the ellipticity and hence the damping is not of the Gilbert form; the 
relaxation time is an algebraic function of the frequency (see appendix). 
\ The damping in the film still shows oscillations as a function 
of the thickness of the film and it attains larger values than in the previous
calculations to first order in $J^2$ 
(fig. \ref{damp6}-\ref{damp7}).

\ The expression for the noise, Eq. \ref{noisexx}, shows a very 
interesting property that  only the Green functions 
$\overline{G}_4$ and $\overline{G}_5$ show up as multiples of 
the coupling constant $J^2$ in the numerator. \ The 
dissipative Green functions 
do not appear in this form. \ This observation is important 
when we try to include the effect of 
spin accumulation on the noise and understand why
 the effective temperature concept appears in the 
stochastic formulation of Li and Zhang \cite{li} for 
close to equilibrium.
\ It is also important 
to observe that in the half-metallic limit, i.e., $ J \rightarrow \infty$, we 
have 
\begin{equation}
\alpha = \frac{\pi\left(  c_{1}^{2}+c_{2}^{2}\right)  }{2^{6}\left(
k_{F}D\right)  } ,
\end{equation}
Hence, the exchange energy splitting drops out completely from the 
damping expression. This result is 
similar to that derived by Bazaliy, Jones and Zhang 
 \cite{bazaliy} in a half-infinite ferromagnet in contact with a normal 
metal.  
\ In the adiabatic limit, $\omega << J$,the expression for the noise can be 
simplified considerably. \ At 
 high temperature (i.e., $k_BT >> w_0$ and $T << T_c$ the critical temperature), it becomes

\begin{equation}
C_{xx}\left(  \omega\right)  =\frac{2\alpha k_{B}T}{\left(c_{1}^2+c_{2}^2\right)}\left(
\frac{ \omega^{2} + c_{yy} + \left(\alpha\omega\right)^{2} }{\left[  \left(  \omega-\omega_{0}\right)  ^{2}+\left(
\alpha\omega\right)  ^{2}\right]  \left[  \left(  \omega+\omega_{0}\right)
^{2}+\left(  \alpha\omega\right)  ^{2}\right]  }\right),
\label{noise2}
\end{equation}
where the damping $\alpha$ is thickness dependent and vanishes when 
$D \rightarrow \infty$ in the film. \ The shift in frequency is less than one 
percent and is neglected in the macroscopic 
case. \ A similar result  derived in the isotropic case has been 
recently communicated to us by \onlinecite{foros}. \ A comparison of this expression with the corresponding expression 
in ref. \onlinecite{smith} reveals a very interesting result; 
 the dynamic of the 
d-electrons in the presence of the sd-exchange interaction 
is simply reproducible
by the GB  equation with white noise and $\alpha$ is replaced by $\alpha_G \left( c_{1}^2+c_{2}^2     \right)$ where $\alpha_G$ is the
Gilbert damping. \ 
\ This is an important simple  result which shows
 the reasons behind the successes
 of the GB equation.\cite{smith} \ At the same time this latter result points 
to the limits  
of applicability of the GB  equation in atomic simulations.\cite{chantrell}
\ The recent 
work of Safonov and Bertram \cite{safonov2} suggests changing 
the damping form of the GB equation. \ Our calculation clearly 
shows that for the average classical magnetization
 this is not needed. \ It is only for high frequencies that 
the effect of the symmetry of the Hamiltonian on the 
damping becomes appreciable (see appendix). \
 For low frequencies, the dependence of the relaxation time on the
ellipticity factor, $c_1^2+c_2^2$,  agrees with
 Kambersky and Patton\cite{kambersky}. \ In this 
work, we are also able to give explicit expressions 
for the damping term and the corresponding noise in thin films. \
  In the next section, 
we will use the GB equation with a damping $\alpha = 0.03$ to 
study the continuum case. \ We will also include 
polarization due to another magnetic layer with relative angle 
far from zero or $180^o$.
\ Next we treat the effect of spin accumulation on the 
noise in the low frequency regime and close to the 
FMR frequency. \  As we stated 
after Eq. \ref{f2f1}, the spin accumulation term will have to 
be included in the self-energy
term. \ The only Green functions where this has 
to be taken into account are those that 
appear as a result of the fluctuations $z$ and $z^\ast$. \ These 
Green functions will affect only terms where the parameter 
$\alpha$ and the frequency $\omega$ appear in the Gilbert 
form in the $\overline G_1$ Green function {\it only}.
\ The overall $\alpha$-term that shows up in front of the 
temperature will therefore remain unaffected by the 
spin accumulation term.  \ To show this result, requires 
a more careful evaluation of the $\overline G_1
(\omega)$. \ The $\mathbf k$-integrals over $\mathbf{k}$ and 
$\mathbf{k}-\mathbf{p}$ are carried out 
separately. \ In the limit of small damping and close 
to the FMR frequency, 
we can define  an effective temperature in the GB 
equation 
\begin{equation}
T^{\ast} = T \frac{1}{\left(1+ \frac{\Delta\mu^{\uparrow
\downarrow}}{\omega_0} \right)},
\end{equation}
and a renormalized interfacial damping 
\begin{equation}
\alpha^\ast = \alpha \left( 1 + \frac{ \Delta\mu^{\uparrow
\downarrow}}{\omega_0} \right).
\end{equation}
This will allow us to write the corresponding noise in a 
form that is remarkably similar to that derived 
recently by Li and Zhang\cite{li} where they considered 
only bulk damping in their 
problem. \ However there is a very important difference. \ The difference 
in chemical potential is expected to be proportional to the 
current\cite{berger,stiles} but not linear to the damping as 
in \onlinecite{li} which is based on the Slonczweski 
picture. \ Therefore this expression should be able to 
differentiate between the sd mechanism and the Slonczewski 
mechanism. \ There are already indirect indication from 
ref. \onlinecite{ryan} that the critical current is not 
linear with damping which is the result
derived in \onlinecite{li} and includes only 
 bulk damping. \ In CPP structures 
therefore we should not expect the spin momentum transfer
term to change the character of noise in these 
systems as long as we stay below critical 
currents. \ There are however other 
effects in this geometry
 not addressed by this calculation. \ One of them 
is the field from the current and the other is 
magnon-phonon 
interactions which my may affect the average 
magnetization configuration in these films.\ We 
discuss these two effects 
classically in the following 
section.

\section{\bigskip discussion and conclusion}

\begin{figure}[ht]
  \begin{center}
 \mbox{\epsfig{file=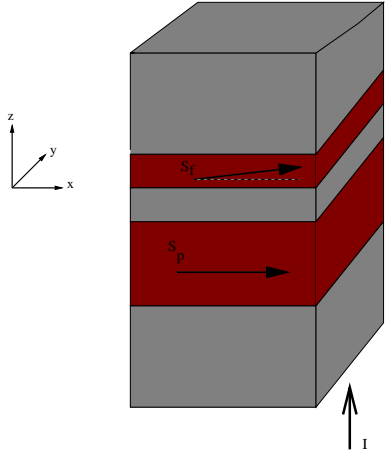,height=6 cm}}
  \end{center}
  \caption{{{The CPP spin valve: The thick layer with magnetization $S_p$ 
is pinned along the x-axis. The magnetization $S_f$ is free to 
move. The current I is perpendicular to the interfaces. Spin momentum 
is transferred from the pinned layer to the layer by 
polarizing the current with the fixed layer.  }}}
  \label{cpp}
\end{figure}

\begin{figure}[ht]
  \begin{center}
 \mbox{\epsfig{file=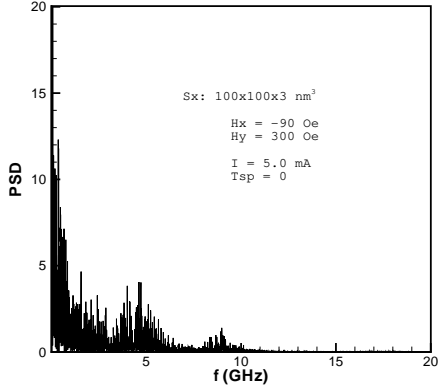,height=6 cm}}
  \end{center}
  \caption{{{The xx-component of the noise spectrum for the average 
$x-$component of the magnetization with the spin torque $T_{sp}$
 between the magnetic layers set to zero in the GB equation.  }}}
\label{sx0}
\end{figure}

\begin{figure}[ht]
  \begin{center}
 \mbox{\epsfig{file=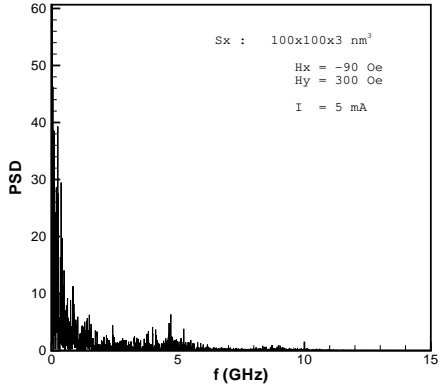,height=6 cm}}
  \end{center}
  \caption{{{ Same as in Fig. \ref{sx0} but with the spin momentum transfer
torque included in the GB equation. }}}
\label{sxaj}
\end{figure}

In this final section, we would like to make some comments and discuss
our results in light of recent experiments on noise 
in spin valves, such as that in ref. \onlinecite{covington},
which show 1/f-type noise in CPP structure . \  In 
CPP 
spin valves with biased fields, it was observed that excessive low 
frequency noise is generated with current. \ It was argued that 
spin momentum transfer between the magnetic layers is responsible 
for this noise. \ In this section, we treat a case 
where the spin momentum transfer is not the root cause 
of noise in these geometries but it will just affect the 
amplitudes of the noise.\ The noise in our case 
will be inherently due to at least three 
contributing factors:
the biasing of the spin valve, the 
field from the current and the thermal fluctuations in the 
system. \ We reach 
these conclusions based on our calculations carried out in previous 
sections 
 and on simulations based on 
the GB equation with a spin torque as it was suggested in 
ref. \onlinecite{slon},

\begin{equation}
 \frac{d \mathbf{S}}{dt}= \gamma \mathbf{S} \times \left( 
\mathbf{H}_{eff} + \alpha_G \frac{\mathbf{dS}}{dt} + \mathbf{h}\right) + \beta I \mathbf{S}
\times \left( \mathbf{S} \times \mathbf{S}_p \right),
\end{equation} 
with the stochastic field $\mathbf{h} ( t ) $ satisfying
\begin{equation}
\langle h_i(t) h_j(t') \rangle = 2 \alpha_G k_B T \delta_{ij} \delta (t-t').
\end{equation}
\ The parameter $\beta$ is a geometrical factor and $I$ is 
the current. \ The pinned layer with magnetization $\mathbf{S}_p$ is not dynamical and 
hence the finite size effects discussed above can be included by choosing 
a large $\alpha_G$. \ The effect of spin 
accumulation is not taken into account 
properly in the stochastic 
field; we are simply assuming the 
effective $\alpha_G$ to be a constant and 
current independent. \ This equation is solved numerically 
with $\alpha_G = 0.02$, $M_s=1400 \; emu/cc$ and thickness $d= 3 \; nm$. \cite{parker,rebei2} \ The white noise approximation as discussed above is  valid  for frequencies around the FMR frequency which is of 
the order of $10 \; GHz$. 
\ There are two theories
 of spin momentum transfer: one is microscopic and 
based on the sd-exchange model while the second is macroscopic 
and is based on a simple balance equation for the spin currents. \ The 
calculations presented in previous sections 
are closer to the first approach rather than to the 
second one. \ It is believed that the first approach is dominant only 
at very thin films of $10 \; nm$ or less.\cite{bauer} \ Therefore based on the results 
derived here, it appears  that any 1/f-type noise 
measured in CPP spin valves should be attributed to non sd-type of 
interactions. \ We show below that the spin torque does not appear to 
 generate 1/f-type noise and it is the non-homogeneities in the 
magnetic configurations that are mainly responsible for the noise. \ 
The CPP structure we study is shown in fig. \ref{cpp} and 
is similar to that in \onlinecite{covington}. \ The current 
is flowing from the pinned to the free layer which are separated by a
normal conducting layer. \ The single particle simulations 
of the magnetization show no interesting behavior and are 
noiseless and this is consistent with the 
results from the previous section. \ The 
field from the current is taken into account in the calculations and 
is needed to observe 1/f-type noise at finite temperature. \ The 
magnetization is also 
biased in the y-axis and x-axis with a $300 \; Oe$ field 
and a $-90 \; Oe$ field, respectively. \ The 
demagnetization field of the 
sample is also taken into account. \ The magnetic material is chosen 
to be that of a permalloy. \ Figures \ref{sx0} and \ref{sxaj} clearly show that 
the effect of the spin torque  only  slightly increases 
the already present noise
in the system for the particular 
parameters shown in the figure. \ It 
{\it does not} give rise to the low frequency in this 
example. \ A closer study of 
 this example shows that it is the combination 
of the biasing, the field from the current and the temperature that are the 
source of the noise. \ Hence in this example, neither the sd-type model nor the 
macroscopic model can explain the origin of the 1/f-type noise. \ A calculation 
that does not include in great detail the configuration of the magnetization 
is therefore highly 
unlikely to capture the source of the noise in these 
structures.\ Figures \ref{bccr04} and \ref{bccr05} show the two 
possible 
metastable states that are responsible for the 1/f-type noise in this 
device.

\begin{figure}[ht]
  \begin{center}
 \mbox{\epsfig{file=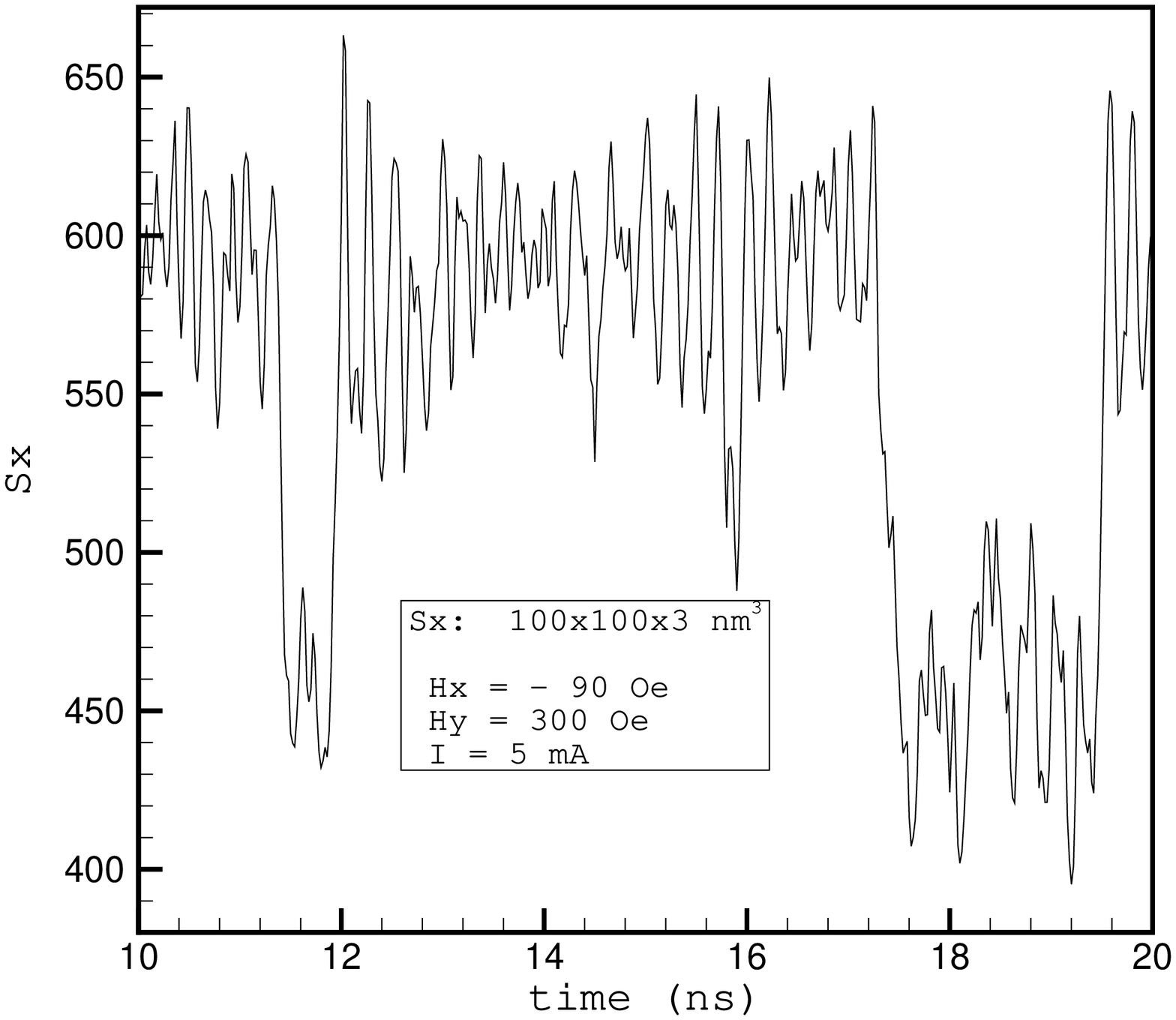,height=6 cm}}
  \end{center}
  \caption{{{A real-time trace of the average magnetization component 
$S_x$ (only a typical interval of time is shown). The magnetization appears to oscillate 
largely in this component for this {\it magnetic} configuration where the total magnetization is mainly along the $y$-axis that is perpendicular to the polarization axis. }}}
\label{sxt}
\end{figure}

\begin{figure}[ht]
  \begin{center}
 \mbox{\epsfig{file=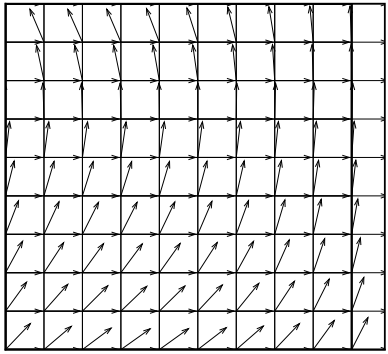,height=6 cm}}
  \end{center}
  \caption{{{Configuration of the magnetization in state 1. The horizontal 
arrows are  along $\mathbf{S}_p$. \ The angle between the external 
bias field and $\mathbf{S}_p$ is close to $90$ degrees.  }}}
\label{bccr04}
\end{figure}

\begin{figure}[ht]
  \begin{center}
 \mbox{\epsfig{file=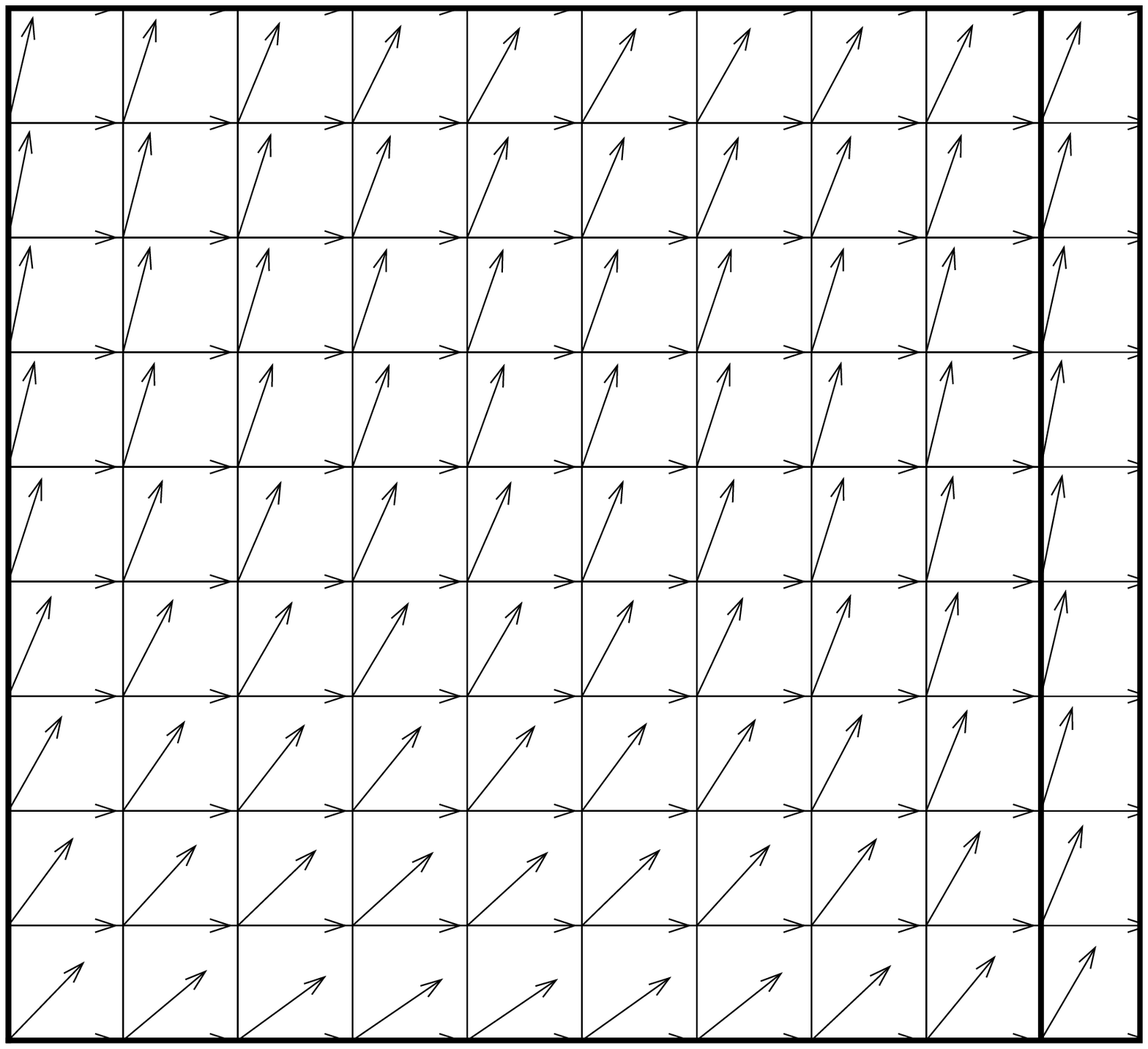,height=6 cm}}
  \end{center}
  \caption{{{Configuration of the magnetization in state 2 (same as in fig. \ref{bccr04}).    }}}
\label{bccr05}
\end{figure}

\bigskip

\ In summary, we have studied in some
detail the noise and the damping problem in thin 
magnetic films embedded between two normal conductors.\ We have 
mainly focused on the interaction between the 
conduction electrons and the d-electrons as 
the main mechanism for damping. \ Our results also 
apply to the microscopic case at high frequencies.\ In 
this model, it has been shown that the damping in thin layers
 oscillates as a function of the thickness of the film. \ If 
higher orders in the exchange coupling constant are taken into account, the relaxation time
becomes dependent on the ellipticity of the precession of the magnetization. \ Only at high frequencies, the damping is no longer 
of the Gilbert form and becomes explicitly dependent 
on the ellipticity.\ This result is important for 
atomic simulations. \
The noise associated with the interfacial 
damping has also been calculated. \ It was shown that it
does not give rise to a large 1/f-type noise.\ The 
spectral density curve gives 
the usual peak at the natural frequency of the 
system at low damping. \ In the adiabatic 
limit, we have found that for small damping 
and close to equilibrium, the 
spectrum is well represented by a white noise source term and an 
effective temperature that measures deviations from 
 small spin accumulations. \ Therefore a large negative 
current can give rise to a large renormalized 
damping and suppression of 
the noise amplitude around the FMR peak while positive current
 decreases the effective damping and increases 
the noise around the FMR peak.\ Finally, we have also shown 
that even within a macroscopic 
calculation the transfer of spin momentum does not 
give rise to large low frequency noise, rather in our example it was the field 
from the current, the thermal fluctuations and a particular biasing that 
simultaneously
give 
rise to large noise at frequencies below the FMR frequency.

\bigskip

\section*{acknowledgment}

The first author  expresses his warmest thanks  to 
G. Bauer, L. Berger, D. Boyanovsky, M. Covington, S. Mukherjee, 
C. Patton, 
E. Rossi, E. Simanek, M. Stiles, V. Safonov, Y. Tserkovnyak and 
S. Zhang for useful discussions that helped shape the text in 
the present form. \  We are also very grateful to  D. L.  Mills 
for sending us  details about his method 
of solution used in his paper.\ (AR) is also very
grateful to Dr. R. Chantrell for his encouragement  and 
support to this work. \
 P. Asselin has kindly commented on 
 parts of the manuscript. 

\appendix

\bigskip

\section{Appendix}

Here we give the definitions of the different functions that appear in the main
text and briefly discuss the high frequency regime of the 
damping term which becomes dependent on the symmetry of 
the Hamiltonian in a non-trivial way.
 \ The following function are derived by integrating out 
the conduction electrons degrees of freedom 
from the magnetization-conduction
electron equations of motion. \ The functions are
\begin{align}
G_{1}\left(  t-t^{\prime},\mathbf{p}\right)   &  =\sum_{\mathbf{k}}\left\{
c_{1}^{2}\mathcal{U}_{12}\left(  t-t^{\prime}\right)  e^{-i\varepsilon
_{12}\left(  \mathbf{k}+\mathbf{p}\right)  \left(  t-t^{\prime}\right)
}\right. \\
&  +\left.  c_{2}^{2}\mathcal{U}_{21}\left(  t^{\prime}-t\right)
e^{i\varepsilon_{12}\left(  \mathbf{k}-\mathbf{p}\right)  \left(  t-t^{\prime
}\right)  }\right\}, \nonumber
\end{align}%
\begin{align}
G_{2}\left(  t-t^{\prime},\mathbf{p}\right)   &  =\sum_{\mathbf{k}}\left\{
c_{1}c_{2}\mathcal{U}_{12}\left(  t-t^{\prime}\right)  e^{-i\varepsilon
_{12}\left(  \mathbf{k}+\mathbf{p}\right)  \left(  t-t^{\prime}\right)
}\right. \\
&  +\left.  c_{1}c_{2}\mathcal{U}_{21}\left(  t^{\prime}-t\right)
e^{i\varepsilon_{12}\left(  \mathbf{k}-\mathbf{p}\right)  \left(  t-t^{\prime
}\right)  }\right\}, \nonumber
\end{align}
and 
\begin{align}
G_{3}\left(  t-t^{\prime},\mathbf{p}\right)   &  =\sum_{\mathbf{k}}\left\{
c_{2}^{2}\mathcal{U}_{12}\left(  t-t^{\prime}\right)  e^{-i\varepsilon
_{12}\left(  \mathbf{k}+\mathbf{p}\right)  \left(  t-t^{\prime}\right)
}\right. \\
&  +\left.  c_{1}^{2}\mathcal{U}_{21}\left(  t^{\prime}-t\right)
e^{i\varepsilon_{12}\left(  \mathbf{k}-\mathbf{p}\right)  \left(  t-t^{\prime
}\right)  }\right\} \nonumber
\end{align}
are due to the correlation terms between the two branches of the CTP 
path of the path integral. The following functions 
are symmetric in time and will be the origin of the correlation 
functions of the random field:
\begin{align}
G_{4}\left(  t-t^{\prime},\mathbf{p}\right)   &  =\sum_{\mathbf{k}}\left\{
c_{1}^{2}\mathcal{U}_{22}\left(  \mathbf{k}+\mathbf{p}\right)
e^{-i\varepsilon_{12}\left(  \mathbf{k}+\mathbf{p}\right)  \left(
t-t^{\prime}\right)  }\right. \\
&  +\left.  c_{2}^{2}\mathcal{U}_{22}\left(  \mathbf{k}-\mathbf{p}\right)
e^{i\varepsilon_{12}\left(  \mathbf{k}-\mathbf{p}\right)  \left(  t-t^{\prime
}\right)  }\right\}, \nonumber
\end{align}

\begin{align}
G_{5}\left(  t-t^{\prime},\mathbf{p}\right)   &  =\sum_{\mathbf{k}}\left\{
c_{1}c_{2}\mathcal{U}_{22}\left(  \mathbf{k}+\mathbf{p}\right)
e^{-i\varepsilon_{12}\left(  \mathbf{k}+\mathbf{p}\right)  \left(
t-t^{\prime}\right)  }\right.  \\
&  +\left.  c_{1}c_{2}\mathcal{U}_{22}\left(  \mathbf{k}-\mathbf{p}\right)
e^{i\varepsilon_{12}\left(  \mathbf{k}-\mathbf{p}\right)  \left(  t-t^{\prime
}\right)  }\right\}.  \nonumber
\end{align}
All these functions are not independent. For example, we have
\begin{equation}
G_{3}\left(  t-t^{\prime},-\mathbf{p}\right)  =-G_{1}\left(  t-t^{\prime
},\mathbf{p}\right).
\end{equation}

In most of our calculations, the 
functions $\mathcal U_{22}$ are approximated by the following 
expressions:

\begin{equation}
\mathcal{U}_{22}\left(  \mathbf{k}+\mathbf{p}\right)  =\frac{1}{2}%
\delta\left(  \varepsilon \left(  \mathbf{k}\right)  -\overline\varepsilon
_{F}\right)  \left(  \frac{K\cdot\mathbf{p}}{m}+\Delta\right)
\coth\left(  \frac{\beta}{2}\right)  \left(  \frac{K\cdot\mathbf{p}}{m}%
+\Delta\right),
\end{equation}%
and
\begin{equation}
\mathcal{U}_{22}\left(  \mathbf{k}-\mathbf{p}\right)  =\frac{1}{2}%
\delta\left(  \varepsilon\left(  \mathbf{k}\right)  -\overline\varepsilon
_{F}\right)  \left(  -\frac{K\cdot\mathbf{p}}{m}+\Delta\right)
\coth\left(  \frac{\beta}{2}\right)  \left(  -\frac{K\cdot\mathbf{p}}%
{m}+\Delta\right).
\end{equation}
These approximations are not necessary but makes the algebra less involved.

The function $G_{1}$ has the exact explicit expression:

\begin{align}
G_{1}\left(  t-t^{\prime},\mathbf{p}\right)   &  =\sum_{\mathbf{k}}\left\{
-c_{1}^{2}\Theta\left(  t-t^{\prime}\right)  \left[  f_{1}\left(
\mathbf{k}\right)  -f_{2}\left(  \mathbf{k}+\mathbf{p}\right)  \right]
e^{-i\varepsilon_{12}\left(  \mathbf{k}+\mathbf{p}\right)  \left(
t-t^{\prime}\right)  }\right.  \\
&  \left.  +c_{2}^{2}\Theta\left(  t^{\prime}-t\right)  \left[  f_{1}\left(
\mathbf{k}\right)  -f_{2}\left(  \mathbf{k}+\mathbf{p}\right)  \right]
e^{i\varepsilon_{12}\left(  \mathbf{k}-\mathbf{p}\right)  \left(  t-t^{\prime
}\right)  }\right\}.  \nonumber
\end{align}

Similar expressions for $G_{2}$ and $G_{3}$ can be easily deduced from that 
of $G_{1}$.

\ All these functions are needed for the calculation of the 
propagators for the fields $Z, Z^{\ast}, z$ and $z^{\ast}$, which are to first 
order in $J^2$
given, respectively by 

\begin{align}
\mathfrak{g}_{1}\left(  t-t^{\prime}\right)   &  =\Theta\left(  t-t^{\prime
}\right)  e^{i\omega_{0}\left(  t-t^{\prime}\right)  }\\
&  +\left(  \frac{J}{2}\right)  ^{2}\int dt^{^{\prime\prime}}\int
d\tau\Theta\left(  t^{^{\prime\prime}}-t\right)  e^{i\omega_{0}\left(
t-t^{^{\prime\prime}}\right)  }\nonumber\\
&  \times\overline{G}_{1}\left(  t^{^{\prime\prime}}-\tau\right)
\Theta\left(  t^{^{\prime}}-\tau\right)  e^{i\omega_{0}\left(  \tau
-t^{^{\prime}}\right)  }+ ...,\nonumber
\end{align}
\begin{align}
\mathfrak{g}_{3}\left(  t-t^{\prime}\right)   &  =-\Theta\left(  t^{\prime
}-t\right)  e^{-i\omega_{0}\left(  t-t^{\prime}\right)  }\\
&  -\left(  \frac{J}{2}\right)  ^{2}\int dt^{^{\prime\prime}}\int
d\tau\Theta\left(  t^{^{\prime\prime}}-t\right)  e^{-i\omega_{0}\left(
t-t^{^{\prime\prime}}\right)  }\nonumber\\
&  \times\overline{G}_{3}\left(  t^{^{\prime\prime}}-\tau\right)
\Theta\left(  t^{^{\prime}}-\tau\right)  e^{-i\omega_{0}\left(  \tau
-t^{^{\prime}}\right)  }+ ... ,\nonumber
\end{align}%
\begin{align}
\mathfrak{g}_{1}^{^{\prime}}\left(  t-t^{\prime}\right)   &  =\Theta\left(
t-t^{\prime}\right)  e^{-i\omega_{0}\left(  t-t^{\prime}\right)  }\\
&  -\left(  \frac{J}{2}\right)  ^{2}\int dt^{^{\prime\prime}}\int
d\tau\Theta\left(  t-t^{^{\prime\prime}}\right)  e^{-i\omega_{0}\left(
t-t^{^{\prime\prime}}\right)  }\nonumber\\
&  \times\overline{G}_{1}\left(  t^{^{\prime\prime}}-\tau\right)
\Theta\left(  \tau-t^{^{\prime}}\right)  e^{-i\omega_{0}\left(  \tau
-t^{^{\prime}}\right)  }+ ... ,\nonumber
\end{align}
and
\begin{align}
\mathfrak{g}_{3}^{^{\prime}}\left(  t-t^{\prime}\right)   &  =\Theta\left(
t-t^{\prime}\right)  e^{i\omega_{0}\left(  t-t^{\prime}\right)  }\\
&  +\left(  \frac{J}{2}\right)  ^{2}\int dt^{^{\prime\prime}}\int
d\tau\Theta\left(  t-t^{^{\prime\prime}}\right)  e^{i\omega_{0}\left(
t-t^{^{\prime\prime}}\right)  }\nonumber\\
&  \times\overline{G}_{3}\left(  t^{^{\prime\prime}}-\tau\right)
\Theta\left(  \tau-t^{^{\prime}}\right)  e^{i\omega_{0}\left(  \tau
-t^{^{\prime}}\right)  }+ ... .\nonumber
\end{align}

\bigskip The various Green functions needed for the calculation of the noise
spectrum are given here in Fourier space $\left(  \hbar=1\right)  .$ The free
propagators associated with the fields $Z(t)$ and $z(t)$ are %

\begin{align}
\mathfrak{g}_{1}^{\left(  0\right)  }\left(  \omega\right)    & =\frac{i}{\omega
-\omega_{0}+i\eta},\\
\mathfrak{g}_{1}^{^{\prime(0)}}\left(  \omega\right)    & =\frac{i}{\omega-\omega
_{0}-i\eta}.
\end{align}
In the presence of the conduction electrons they become $\left(
\lambda=J/2\right)  $,

\begin{equation}
\mathfrak{g}_{1}\left(  \omega\right)  =\frac{\mathfrak{g}_{1}^{\left(  0\right)  }\left(
\omega\right)  }{1-\lambda^{2}\overline{G}_{1}\left(  \omega\right)
\mathfrak{g}_{1}^{\left(  0\right)  }\left(  \omega\right)  }
\end{equation}
and
\begin{equation}
\mathfrak{g}_{1}^{^{\prime}}\left(  \omega\right)  =\frac{\mathfrak{g}_{1}^{^{\prime}\left(
0\right)  }\left(  \omega\right)  }{1+\lambda^{2}\overline{G}_{1}\left(
\omega\right)  \mathfrak{g}_{1}^{\prime\left(  0\right)  }\left(  \omega\right)  }.
\end{equation}
We will also need their conjugate form, which is again different from the
complex conjugate. The free propagator is%
\begin{equation}
\mathfrak{g}_{1}^{^{\prime\ast(0)}}\left(  \omega\right)  =\frac{i}{\omega+\omega
_{0}+i\eta},
\end{equation}
and the dressed one is
\begin{equation}
\mathfrak{g}_{1}^{^{\prime\ast}}\left(  \omega\right)  =\frac{\mathfrak{g}_{1}^{^{\prime}\ast\left(
0\right)  }\left(  \omega\right)  }{1-\lambda^{2}\overline{G}_{3}\left(  \omega\right)  \mathfrak{g}_{1}^{^{\prime\ast}\left(  0\right)  }\left(
\omega\right)  }.
\end{equation}
The kernels $\overline{G}_{\alpha}$, $\left(  \alpha=1,4,5\right)  $, are
averaged over all spin wave modes. For $\overline{G}_{1}$, it is defined by
\begin{align}
\overline{G}_{1}\left(  \omega\right)    & = 
D \int\frac{d^{3}\mathbf{k}}{\left(
2\pi\right)}\int_{-\infty}^{+\infty}\frac
{dp}{2\pi}\left|  \frac{\sin\left(  p\frac{D}{2}\right)  }{p\frac{D}{2}%
}\right|  ^{2}\\
& \times\left\{  c_{1}^{2}\frac{\partial f}{\partial\varepsilon_{k}}\left(  \frac{k_{x}p}{m}+\delta\right)  \frac{i}{\delta+\frac{k_{x}p}%
{m}-\omega+i\eta}\right.  \nonumber\\
& +\left.  c_{2}^{2}\frac{\partial f}{\partial\varepsilon_{k}
}\left(  \frac{-k_{x}p}{m}+\delta\right)  \frac{i}{\delta-\frac{k_{x}p}%
{m}+\omega-i\eta}\right\}  ,\nonumber
\end{align}
with similar definitions for the other kernels. The Fourier transforms are
approximately given by
\begin{align}
{\Im} \; \overline{G}_{1}\left(  \omega\right)    & = v m D\int_{0}^{\infty}\frac{dp}{2\pi
}\left|  \frac{\sin\left(  p\frac{D}{2}\right)  }{p\frac{D}{2}}\right|
^{2}\\
& \times\left\{  c_{1}^{2}\left(  2 \overline k_{F} + \frac{m\omega}{p}
\log\left[\left|  \frac{\delta-\omega+\frac{k_{F}p}{m}}{\delta-\omega-\frac{k_{F}
p}{m}}\right| \right]  \right)  \right.  \nonumber\\
& +\left.  c_{2}^{2}\left(  2 \overline k_{F} +\frac{m\omega}{p}\log\left[ \left|
\frac{\delta+\omega-\frac{k_{F}p}{m}}{\delta+\omega+ 
\frac{k_{F}p}{m}}\right| \right]
\right)  \right\}  ,\nonumber
\end{align}
and
\begin{align}
\overline{G}_{4}\left(  \omega\right)    & =\frac{m^2 v}{2}\omega\coth\left(
\beta\frac{\omega}{2}\right)  D\int_{0}^{\infty}\frac{dp}{\pi p}\left|
\frac{\sin\left(  p\frac{D}{2}\right)  }{p\frac{D}{2}}\right|  ^{2}\\
& \times\left\{  c_{1}^{2}\Theta\left(  \overline k_{F} -\left|
\frac{m\left(  \omega-\delta\right)  }{p}\right|  \right)  +c_{2}^{2}%
\Theta\left(  \overline k_{F} -\left|  \frac{m\left(  \omega+\delta\right)
}{p}\right|  \right)  \right\}  ,\nonumber
\end{align}
and
\begin{align}
\overline{G}_{5}\left(  \omega\right)    & =\frac{m^2 v c_{1}c_{2}}{2}\omega
\coth\left(  \beta\frac{\omega}{2}\right)  D^{2}\int_{0}^{\infty}%
\frac{dp}{\pi p}\left|  \frac{\sin\left(  p\frac{D}{2}\right)  }{p\frac{D}{2}%
}\right|  ^{2}\\
& \times\left\{  \Theta\left(  \overline k_{F}-\left|  \frac{m\left(
\omega+\delta\right)  }{p}\right|  \right)  +\Theta\left(  \overline k_{F} -\left|  \frac{m\left(  \omega-\delta\right)  }{p}\right|  \right)  
\right\}
.\nonumber
\end{align}
\ In all of the above expressions, $\overline k_F = \frac{k_F^{\uparrow}
+ k_F^{\downarrow}}{2}$. \
It should be also noted that 
\begin{equation}
\overline{G}_{1}^{\ast}\left(  \omega\right)  =\overline{G}_{1}\left(
-\omega\right).
\end{equation}

\ Finally we give the term that is directly responsible 
for the relaxation term,
\begin{equation}
\mathcal{R}e\overline{G}_{1}=\frac{\pi mDv}{4}\omega\int\frac{dx}{x}\left(
\frac{\sin x}{x}\right)  ^{2}\left[  c_{1}^{2}\Theta\left(  \frac{2k_{F}}%
{mD}x-\left|  \omega-\Delta\right|  \right)  +c_{2}^{2}\Theta\left(
\frac{2k_{F}}{mD}x-\left|  \omega+\Delta\right|  \right)  \right]
,\label{relaxation}
\end{equation}
where $c_{1}^{2}=\left(  \sqrt{c_{xx}}+\sqrt{c_{yy}}\right)  ^{2}/\left(
4\omega_{0}\right)  $ and $c_{2}^{2}=\left(  \sqrt{c_{xx}}+\sqrt{c_{yy}%
}\right)  ^{2}/\left(  4\omega_{0}\right)  .$ \ For the noise
in the
average 
magnetization, the range of frequencies we are interested 
in are usually
  very low compared to the
exchange splitting energy. \ In this case
  the relaxation time which is proportional to
$\mathcal{R}e\overline{G}_{1}$ will depend only 
on the overall ellipticity
factor $c_{1}^{2}+c_{2}^{2}$.\ The cases of higher 
frequencies are of interest
only in the atomistic limit which will also give a 
similar expression as in
\ref{relaxation} for the relaxation 
time. \ In this case, the relaxation as easily seen will depend
separately on $c_{xx}$ and $c_{yy}$ and is no 
longer linear in $\omega.$ \
Memory effects in the systems become important 
and the GB equation is no
longer valid. \ In the linear model of ref.
\onlinecite{rebei}, we did not have  any higher 
order dependence on frequency
and the damping was of the Gilbert form. \ Hence 
representing a bath by harmonic oscillators is 
only useful when we are interested in the low energy 
limit.

\end{document}